\documentclass{ametsocV6.1}
\title{Understanding Left-Moving Supercells: Environmental Factors and Forecasting Challenges}

\authors{Aaron W. Zeeb\aff{a}\correspondingauthor{Aaron W. Zeeb, Brooks Hall 321, Central Michigan University, Mt. Pleasant, MI 48858, USA. Email: zeeb1a@cmich.edu}, John T. Allen\aff{a}, Matthew Van Den Broeke\aff{b}}

\affiliation{\aff{a}{Department of Earth and Atmospheric Sciences, Central Michigan University, Mt. Pleasant, Michigan, USA}, \aff{b}{Department of Earth and Atmospheric Sciences,  University of Nebraska--Lincoln, Lincoln, Nebraska, USA}}

\abstract{Left-moving (LM) supercells, characterized by anticyclonically rotating updrafts in the Northern Hemisphere, are significant due to their propensity to produce large hail. Although less common than right-moving supercells, they present notable forecasting challenges and societal impacts. However, despite these impacts, the environments of LM supercells are poorly understood compared to their right-moving counterparts. To address this gap, this research focuses on enhancing the understanding of LM supercells by examining the environmental conditions conducive to their development. A manually compiled and quality-controlled dataset of over 850 LM supercell cases across North America is used to provide a robust sample. Near-storm environments are characterized through the use of RAP/RUC inflow proximity sounding profiles. Leveraging storm properties, including mesoanticyclone strength, hail size, wind speed, and duration, we investigate whether environments can differentiate between these varying strengths and categories, thereby enhancing forecaster awareness. Results show that LMs typically form in environments supportive of right movers, with a key difference being that LMs likely only realize the shape of the hodograph above their LCLs. Lapse rates, CAPE, and LCL height are the best predictors of LM strength and hail potential. LMs with wind reports have drier boundary layer moisture, steeper 0--3 km lapse rates, larger CAPE, and higher LCL heights, leading to increased evaporational cooling. Longer-lived LMs often have weaker CAPE and stronger shear as compared to shorter-lived LMs. These results establish a unique parameter space climatology of LM supercells, thus providing essential forecasting insight and reducing the research gap for these storms.
} 

\begin{document}


\maketitle

%
%
%
\statement{
Supercells produce the majority of severe weather perils of any convective storm, posing a significant threat to society. These storms often develop in pairs and move either to the right or left of the mean atmospheric wind (for typical wind profiles), resulting in two distinct storms and associated hazards. The left-moving supercell, while less common than the right-mover, has been studied more rarely even though they are perceived as significant producers of severe hail and wind. This has resulted in a knowledge gap and substantial forecasting challenge. This research establishes the environmental ingredients in which these leftward moving supercells form to aid forecasters in predicting and warning for such events.
}
%
%

%
\section{Introduction}

    Severe convective storms pose a great threat to society, resulting in more than \$60 billion in annual insurance losses \citep{Aon2025, MunichRe2025}. In the United States, right-moving (RM) supercells (i.e., cyclonically rotating and moving to the right of the mean vertical wind shear) are commonly documented for their propensity to produce multiple hazards (e.g., hail, wind, and tornadoes), leading to a high contribution to annual losses. In contrast, their left-moving (LM) counterparts are less studied, mainly due to their perceived uncommon occurrence and predominance of producing hail as opposed to tornadoes, which are typically perceived as a greater threat \citep{brown_evolution_1994, dostalek_aspects_2004, lindsey_observations_2005, bunkers_documentation_2007}. Prior work has approximated that for every LM supercell, there are 50 RMs \citep{burgess1981evidence}; however, this is not observed in operations, and more recent studies have shown LMs to account for 10--30\% of all supercells \citep{bunkers_vertical_2002,homeyer_2025}. Furthermore, while rarer than RM supercells, LMs commonly produce severe hail ($\ge$2.54 cm), with the fraction being as large as 90\% for some climatologies  \citep{bunkers_vertical_2002}. Additionally, some studies have shown that about 75\% of LMs in their dataset produced significant severe hail \citep[$\ge$5.08 cm;][]{edwards_assessment_2004}, which can result in billion-dollar losses in a single event \citep{blair_radar-based_nodate, johnson_evaluation_2014, gunturi_impact_2017}. Although LM supercells pose a substantial threat to society, there is a notable lack of research to understand these severe storms. \par
    The physical processes driving the deviant motion of LM supercells are well understood and arise from the internal dynamics of a rotating updraft. This causes the storm to either split or develop stronger upward motion on the left flank of the updraft, ultimately leading to the most defining characteristic of LM supercells: an anticyclonically rotating updraft \citep[e.g.,][]{weisman_characteristics_1986, klemp_dynamics_1987, davies-jones_linear_nonlinear_propagation_2002}. LMs tend to be weaker, shorter-lived, and rarer than their RM counterparts, at least partly due to the climatological likelihood of veering in the low-level wind field, which in the US typically favors upward vertical motion on the right flank of the updraft \citep{fujita_split_1968, klemp_simulation_1978, davies-jones_streamwise_1984, davies-jones_review_2015, rotunno_rotation_1985, grasso_dissipation_2000}. Furthermore, with few documented LMs producing tornadoes \citep{brown_iowa_1980, monteverdi_first_2001, dostalek_aspects_2004, bunkers_documentation_2007, homeyer_2025}, these events are often seen as less of a hazard by meteorologists. As such, LM supercells have been studied mainly through case studies and radar analyses \citep[e.g.,][]{hammond_study_1967, charba_structure_1971, nielsen-gammon_detection_1995, grasso_observations_2001, edwards_photographic_2006}. Most of these studies focused on documentation of the environmental characteristics of the case and the evolution of the storm, usually through radar and/or satellite analysis, to identify differences relative to their RM counterparts. Several case studies focused on more impactful LM events that produced large hail \citep[e.g.,][]{edwards_assessment_2004, lindsey_observations_2005, zeitler_operational_2005, jones_short-term_2017}, however, none of these studies explicitly explored a large sample of environments that promote the formation of these storms.\par
    Other studies have examined datasets of LMs, but these have been limited to sample sizes of less than a hundred to a few hundred cases, owing to the bias towards focusing on RMs in observations \citep{houze_hailstorms_1993, bunkers_vertical_2002, edwards_assessment_2004, nixon_distinguishing_2022, tonn_evaluating_2023}. These studies found that LM supercells exist in environments supportive of RMs with moderate to strong CAPE, both positive and negative storm-relative helicity (SRH) in the low levels, and a dominant veer-back-veer hodograph shape. Much of the prior literature on LMs focused on the hodograph shape, using it to predict their deviant motion and/or to understand how they acquire their rotation \citep[e.g.,][]{davies-jones_streamwise_1984, bunkers_predicting_2000, bunkers_vertical_2002, Bunkers_2024_motion}. Aside from these metrics, there has been no in-depth research defining the environments that are supportive of LM supercells, especially as they pertain to LM rotational strength. This has translated into fewer tools for operational forecasting, likely compounded by reduced observations from these storms (e.g., storm chasers rarely focus on LM storms).\par
    To address some of these gaps, we build on earlier work by using a new dataset of 889 quality-controlled LM cases occurring from 2011 to 2022 \citep{van_den_broeke_left-moving_2024}. This novel dataset is currently the largest observed and quality-controlled dataset of LMs with quantified mesoanticyclone strength and other metadata for each supercell. Recent studies have begun to grow in sample size \citep[e.g.,][]{nixon_distinguishing_2022, tonn_evaluating_2023, homeyer_2025}, although their datasets have limited quality control and no quantified strength. This research will re-evaluate the current knowledge regarding the environments of LM supercells, exploring their basic thermodynamic and kinematic parameters by focusing on the following questions:

1) What is the typical parameter space in which left-moving supercells develop?

2) What does a typical hodograph and sounding look like for left-moving supercells?

3) How do both of these properties vary based on mesoanticyclone intensity?

\section{Data and Methods}

    This research uses the LM dataset from \citet{van_den_broeke_left-moving_2024}, described further in \citet{Bunkers_2024_motion} and \citet{van_den_broeke_climo_LM_2025}. The reader is referred to \citet{Bunkers_2024_motion} for further details on the dataset and its potential biases. In summary, it consists of 889 observed and quality-controlled LM supercells from 2011 to 2022 with quantified mesoanticyclone strengths. Mesoanticyclone strength—marginal, weak, moderate, and strong—is based on the maximum rotational velocity classifications from the \citet{andra_1997} nomogram. The weakest LMs (i.e., marginal) in this classification closely correspond to the LM members described by \citet{brown_evolution_1994} and may also align with the “intermediary type” supercells identified by \citet{houze_hailstorms_1993}. Spatially, the dataset spans the continental United States, with a concentration across the Great Plains, extending from Texas to the Dakotas (Fig. 1). There is a gradual decrease in cases with eastward extent, and a sharp cutoff west of the Continental Divide, consistent with prior radar-derived supercell climatologies \citep[e.g.,][]{smith_convective_2012, homeyer_2025} and model proxy climatologies \citep{ashley_future_2023, zeeb_supercell_2024}. As radar coverage is not limited by borders, two cases are located just north of North Dakota in Canada (Fig. 1), in addition to four weak cases in Hawaii (not shown). The Hawaiian cases are not included in the data analysis, therefore dropping our total cases to 885. Following radar identification of cases, storm reports from the Storm Prediction Center (SPC) were matched with each LM in the dataset; however, not all cases had reports. All reports were taken within 20 km of each storm centroid during the analysis period and quality controlled by a team of meteorologists to ensure each report was associated with the corresponding storm [see \citet{van_den_broeke_climo_LM_2025} for further details]. The dataset includes a total of 1383 storm reports across 502 cases. Since the reports come from the SPC dataset, reports in the analysis are divided up into their two main categories: severe and significant severe \citep{Hales1988}. Herein, severe hail is $\ge$2.54 cm and $<$5.08 cm, significant severe hail is $\ge$5.08 cm, severe wind is $\ge$50 kts and$<$65 kts, and significant severe wind is $\ge$65 kts.\par
\begin{figure}
    \centering
    \includegraphics[width=1\linewidth]{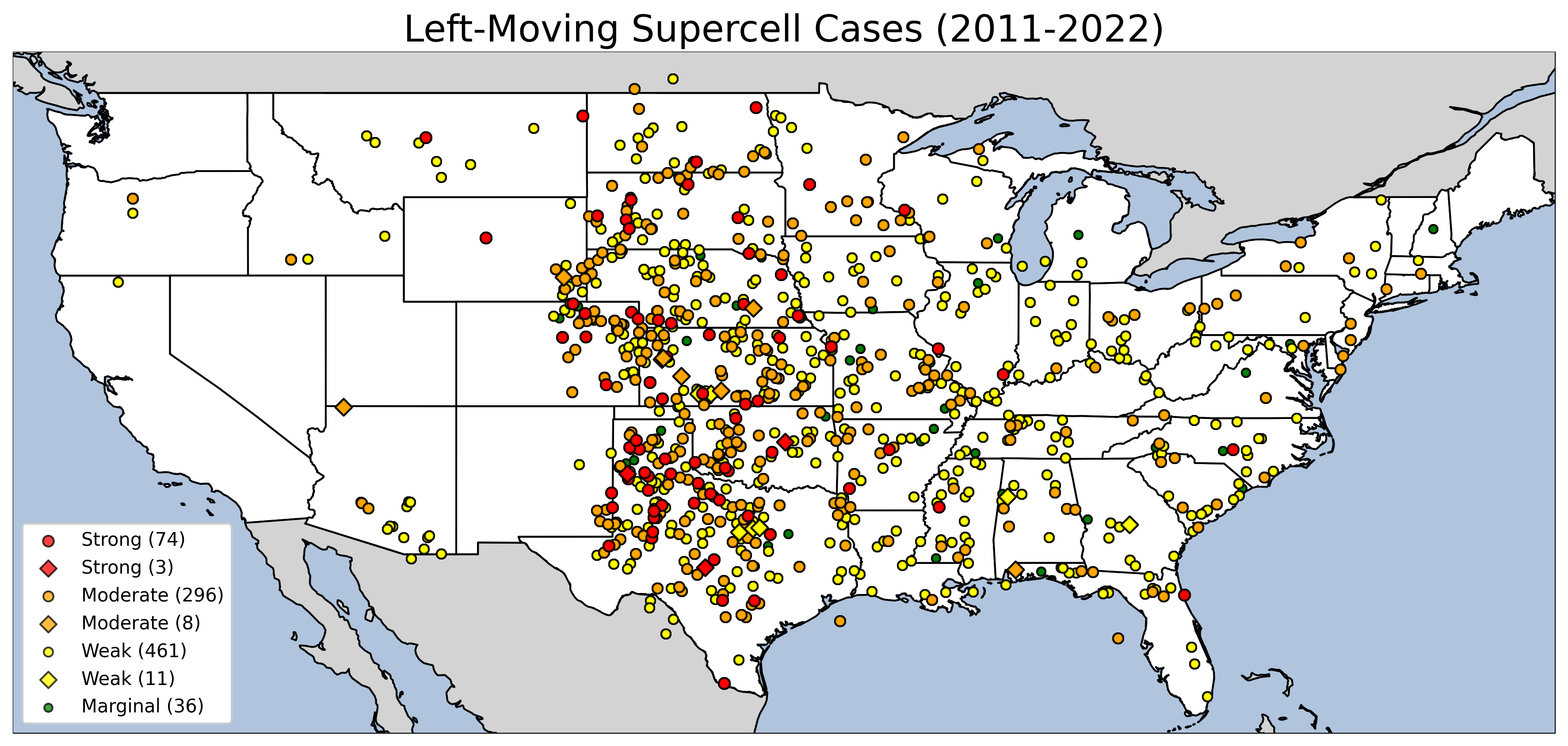}
    \caption{Spatial distributions of the initial latitude and longitude from the 885 LM cases in North America. The colors are assigned by mesoanticyclone intensity and the size of the symbol increases with increasing intensity. Circles correspond to the available 867 RUC/RAP vertical-sounding profiles while the diamonds are the remaining 22 unavailable RAP/RUC profiles.}
    \label{fig:1}
\end{figure}
    As in \citet{Bunkers_2024_motion}, soundings from the Rapid Update Cycle \citep[RUC;][]{Benjamin_2004} and the Rapid Refresh \citep[RAP; operational since May 2012,][]{benjamin_north_2016} models were obtained for each case via the SPC mesoscale analysis system \citep{bothwell_2002}. The nearest inflow sounding at the bottom of the starting hour was selected and derived from post-processed 25-hPa vertical spacing at the nearest model grid point, with surface values based on the RUC/RAP final analysis \citep{coniglio_2022}. The number of vertical data levels varied depending on the surface pressure, with a maximum of 37 levels and a profile top at 100 hPa. These profiles are commonly used for calculating convective parameters and have been shown to reasonably approximate less frequently available rawinsonde profiles, with some limitations \citep{coniglio_verification_2012, coffer_2020,coniglio_2022}. A subset of these data was interpolated to 250-m vertical intervals for skew-T and hodograph construction, while the pressure level data were used for thermodynamic and kinematic parameter calculations. All LM kinematics were calculated using the recently proposed 5 m s$^-$$^1$ deviation off the 0--6 km mean wind as opposed to prior estimates that were symmetrical with the right mover \citep[7.5 m s$^-$$^1$;][]{Bunkers_2024_motion}. All parameters were calculated using the Python package \textit{xcape} \citep{lepore_xcape_2021}. \par
    The dataset includes 867 cases with matching RAP/RUC profiles, as 22 cases have missing profiles due to model run availability. None of the point sounding profiles are convectively contaminated, as assessed by having omega $>$ -10 $\mu$b s$^-$$^1$, $<$90\% mean RH in the lowest 6 km, and $<$0.01 mm of convective precipitation for each grid point. The sample size for the available RAP/RUC soundings for each mesoanticyclone intensity is as follows: 74 strong cases (8.54\%), 296 moderate cases (34.14\%), 461 weak cases (53.17\%), and 36 marginal cases (4.15\%). Results are most robust for moderate and weak cases and hence could potentially skew the results for the “all” (i.e., the entire 867 LM dataset) category; however, since no research has established a climatology for the parameter space of LMs, this has little influence on the impact of this study. During our analysis, we identified another subset of LMs that included elevated cases on the north side of a frontal boundary. These cases are included in the parameter breakdown but separated from the rest of the dataset for the skew-T and CAPE with respect to height analysis, as the substantial temperature inversion makes it uncertain which parcel trace should be used to characterize the profile. \par
\section{Results}

\subsection{Overview}
    LM supercells are most common from March to September (the interquartile range is from April to August), peaking in May with a median of June \citep[Fig. 2a;][]{smith_convective_2012,homeyer_2025}. Marginal and weak LM cases have a median in May, while moderate and strong median occurrences are in June. LM cases in the dataset are most common between 15:00--19:00 local time (LT; 20:00--01:00 UTC) and peak at 17:00 LT, while marginal, weak, and strong cases peak at 22:00 UTC and moderate LM cases peak at 23:00 UTC (Fig. 2b). Generally, LM occurence is two hours earlier compared to RMs \citep{smith_convective_2012} with a median occurrence at 17:00 LT (21:00 UTC), except for strong cases, with a median occurrence at 18:00 LT. Strong cases have the smallest distribution in monthly and LT occurrence, occurring during the seasonal and daily peak in CAPE (Figs. 2a,b). \par
    \begin{figure}
    \centering
    \includegraphics[width=1\linewidth]{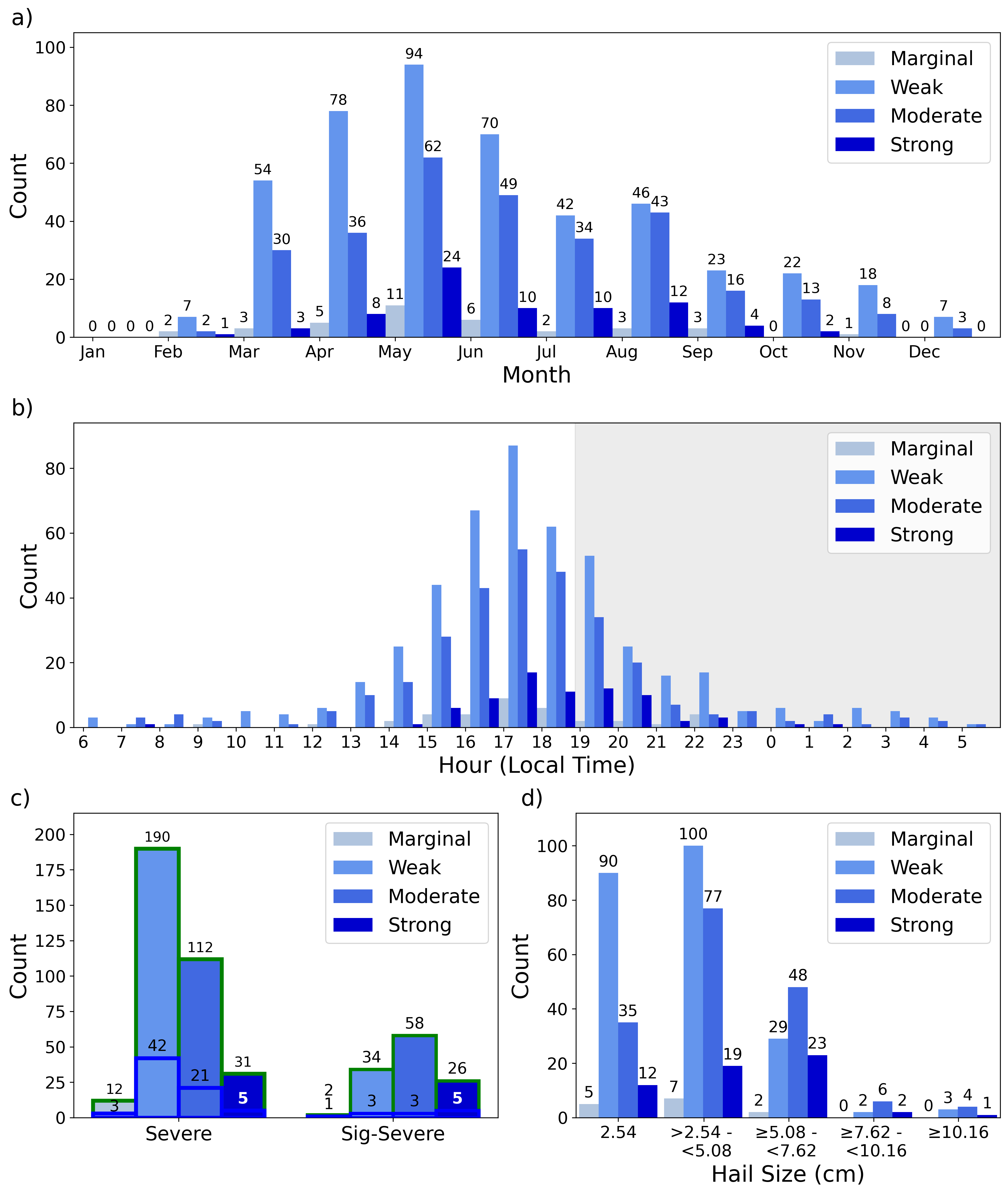}
    \caption{Summary of the month (a) and local time start hour (b) for each case broken down by mesoanticyclone strength. c) and d) are summaries of the maximum magnitude severe weather reports by case for severity (a) and hail size only (b). In c), bars are outlined based on storm report type; green outlines are hail and blue outlines are wind reports.}
    \label{fig:2}
\end{figure}
    While biases in reporting frequency reduce sample size, 502 (56.5 \%) cases have reported hazards. The perception that LM supercells are associated with large hail is supported by this sample, as 1236 (89.4\%) of the storm reports in this dataset are hail reports. Furthermore, 476 (94.8\%) of the cases with reports produced hail $\ge$2.54 cm, 120 (23.9\%) of which produced significant severe hail (Figs. 2c,d). LMs may also produce severe winds, and while not as common as hail reports, there are 145 wind reports (10.5\% of all reports) over 83 (16.5\%) cases. From those cases, 71 cases produced severe wind and 12 cases produced significant severe wind (Fig. 2c). There is only one tornado report from a strong LM. Weak and moderate LMs have the most total severe reports, resulting from their larger sample size in the dataset. \par
\subsection{Thermodynamic Characteristics}
    Increasing LM mesoanticyclone strength shows a positive association with steeper lapse rates (Figs. 3a,d) and results in a significant (\textit{p} $<$ 0.05) relationship between increasing CAPE and mesoanticyclonic intensity, no matter the initial parcel [e.g., surface-based (SB), mixed-layer (ML), most-unstable (MU); Figs. 3b,c]. There are insignificant differences in the CIN distributions across mesoanticyclone strengths, with generally consistent ML CIN distributions and a subtle decreasing interquartile range for MU CIN with increasing mesoanticyclone strength (Figs. 3e,f). Although not significant for all categories, LCL and freezing level heights increase with increasing mesoanticyclone strength (Figs. 3h,g). The median LM LCL height is $\sim$900 m and the median freezing level height is $\sim$3.6 km, both of which are lower than typical RM heights \citep[e.g.,][]{edwards_nationwide_1998, Rasmussen_climatology_sounding_parameters_1998, brooks_climatology_sounding_2002, thompson_close_2003, nixon_hodographs_2023}.\par
    \begin{figure}
    \centering
    \includegraphics[width=1\linewidth]{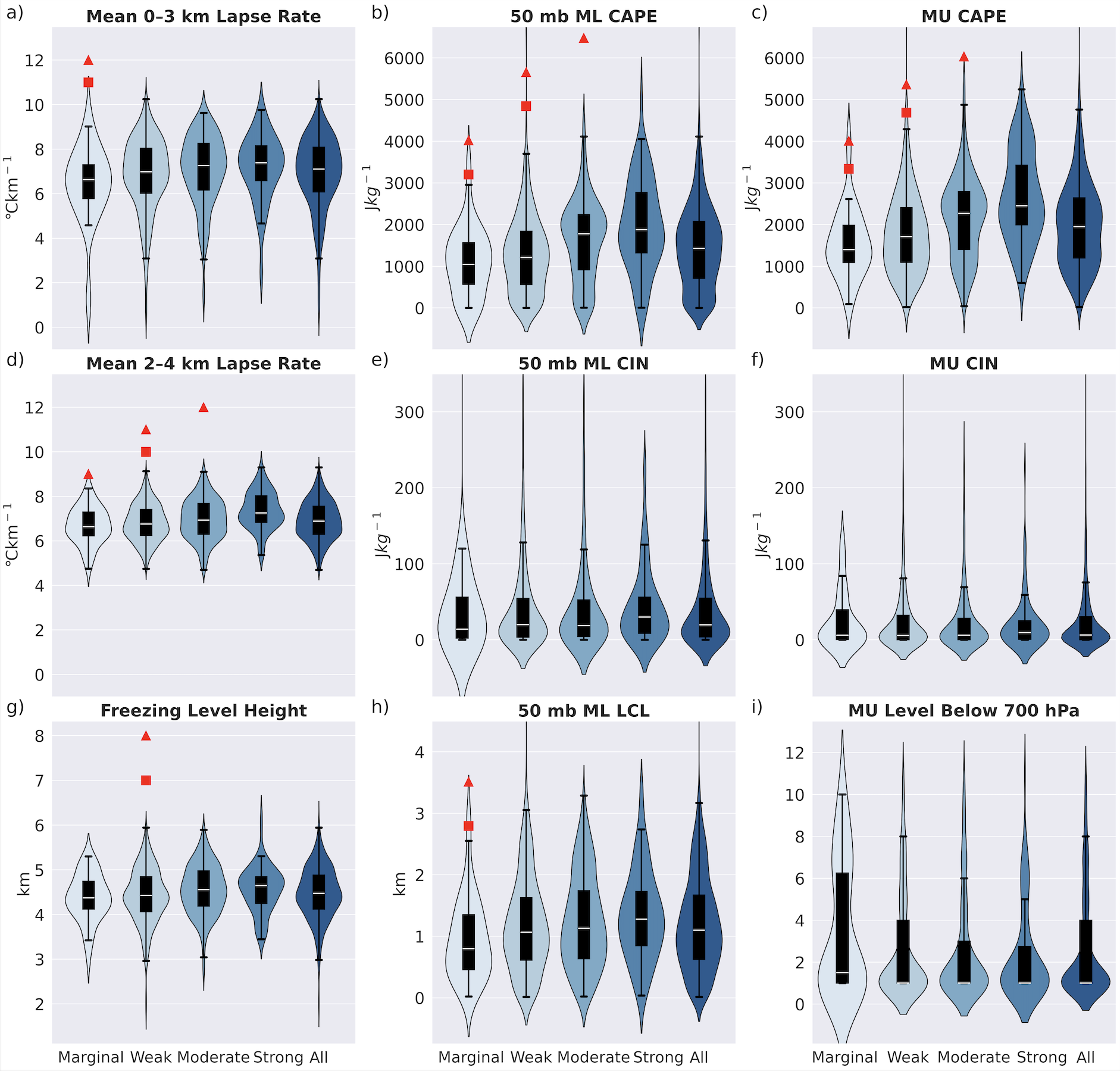}
    \caption{Violin plots of several thermodynamic variables broken down by mesoanticyclone strength for the 867 RUC/RAP sounding profiles in the LM dataset (i.e., the all category). The white line in each box is the median, the box bounds represent the 25\textsuperscript{th} and 75\textsuperscript{th} percentiles, and the whiskers outline the maximum and minimum, excluding outliers. The moderate (strong) mesoanticyclones are statistically different (i.e., \textit{p} $<$ 0.05) compared to the category if a red square (triangle) is drawn using the Brunner-Munzel test. The sample size for each mesoanticyclone intensity is as follows: 36 marginal cases, 461 weak cases, 296 moderate cases, 74 strong cases and 867 total (i.e., all) cases.}
    \label{fig:3}
\end{figure} 
    Since LMs typically move poleward in the Northern Hemisphere, they may be elevated, existing on the cool side of a boundary or transition to an elevated state as they cross over to the cool side of a boundary \citep{bunkers_vertical_2002}. Additionally, since elevated convection is known for producing hail \citep{grant_elevated_1995, banacos_use_2005, horgan_five-year_2007, reif_20-year_2017, nixon_hodographs_2023} , and LMs tend to predominantly generate hail, we investigated how many LMs in the dataset may be elevated and whether the hail producing LMs in the dataset may be elevated. We did not, however, look into cases that may have become elevated after the start of the analysis period when the proximity sounding was obtained, as storm modification likely makes such samples non-representative. The most unstable level below 700 hPa (MU level) indicates whether the most unstable parcel is elevated. Since a storm's inflow cannot always be determined from a modeled sounding, we emphasize that while an MU level of 1 often represents a surface-based profile and an MU level greater than one often represents an elevated profile, this may not correctly categorize the storm's inflow as dynamic processes can result in the lifting of stable air. Nevertheless, the median MU level for all LMs is 1 with the interquartile range decreasing with mesoanticyclone strength, indicating that stronger LMs are more likely surface-based in the nearest sounding profile (Fig. 3i); 50\% of marginal cases have an MU level $>$1. More specifically, the MU level is $>$1 for 41.1\% of all LMs, signifying that these cases may be elevated in the nearest profile (Fig. 3i). Out of these cases, 54.3\% (40.8\% of all hail cases) produced hail; 41.2\% of which was severe hail (41.3\% of all severe hail cases), and 13.2\% significant hail (39.2\% of all significant severe hail cases). \par
    Longer-lived LMs tend to have higher MU levels, decreasing with mesoanticyclone strength (not shown). Of the 59 cases lasting $\ge$2 hours, 32 (54.2\%) have an MU level $>$1, 24 (40.7\%) of which produced severe hail, and 23 (39.0\%) of which produced significant severe hail. Of the 32 cases that have an MU level $>$1, there is an equal number of cases that produced severe and significant severe hail. These results may be biased by the small sample size and the increased reporting coverage that long-lived events create as they travel longer distances. Moreover, the longest-lived LM events are associated with lower CAPE, with median values $<$1000 J kg$^-$$^1$ , whereas shorter-lived LMs have medians $\sim$2000 J kg$^-$$^1$, contrasting with previous work showing CAPE increasing as a function of duration \citep[e.g.,][]{brimelow_1999, brimelow_modeling_2002, jewell_evaluation_2009}. \par
    For LM cases with hail reports, there is a positive association of steeper lapse rates with mesoanticyclone strength and increasing hail size (Figs. 4a,d). This contributes to the positive association of CAPE with increasing hail size and LM strength, where CAPE tends to be slightly higher for significant severe hail compared to severe hail (Figs. 4b,c). Additionally, LMs that produce hail tend to have higher LCLs than those that do not (not shown), with significant severe hail having the highest median LCL heights (Fig. 4h), which is often an important characteristic of hail-producing storms \citep{edwards_nationwide_1998, groenemeijer_sounding-derived_2007, johnson_evaluation_2014, pucik_proximity_2015}. Several previous studies have shown little skill in the freezing level height for predicting hail size or hail potential \citep[e.g.,][]{edwards_nationwide_1998, johnson_evaluation_2014, giordani_characterizing_2024}, although, recently, lower freezing level heights have been shown in hail producing RMs \citep[e.g.,][]{nixon_distinguishing_2022}. For LMs, there are slightly higher freezing levels for significant severe hail cases compared to severe hail \citep[Fig. 4g]{gensini_hail_2021}; however, while statistically significant, the difference is not large and may be difficult to differentiate operationally. \par
\begin{figure}
    \centering
    \includegraphics[width=1\linewidth]{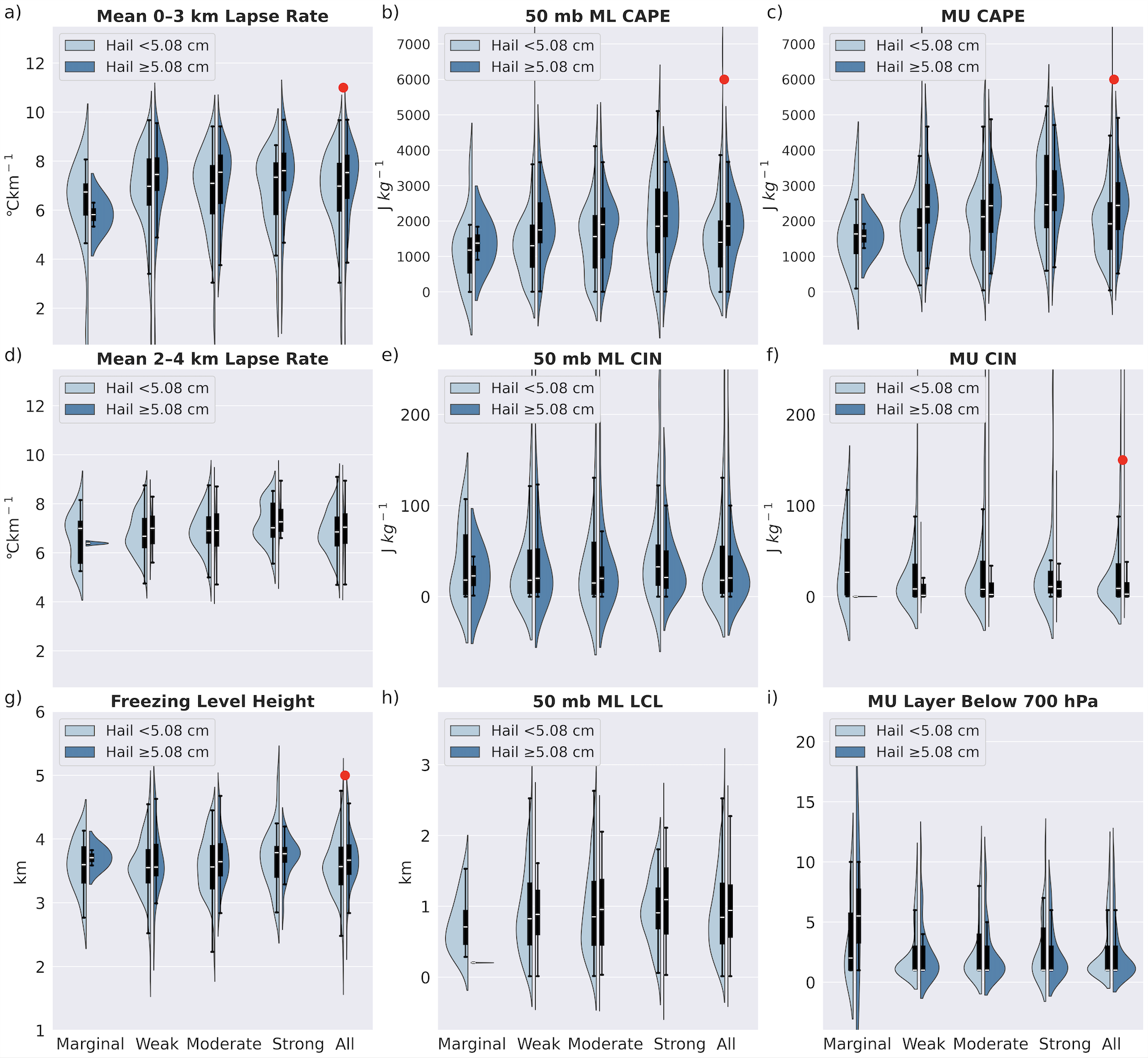}
    \caption{As for Fig. 3, however, only for cases with hail reports. Cases are divided into two groups: severe hail and significant severe hail. A red circle denotes a statistically significant difference (i.e., \textit{p} $<$ 0.05) between severe hail and significant severe hail using the Brunner-Munzel test. The sample size for each mesoanticyclone intensity and hail size is as follows: 12 severe hail and 2 significant severe hail marginal cases, 190 severe and 34 significant severe weak cases, 112 severe and 58 significant severe moderate cases, 31 severe and 26 significant severe strong cases, and 345 severe and 120 significant severe total (i.e., all) cases.}
    \label{fig:4}
\end{figure}
    LM cases with wind reports tend to have steeper 0--3 km lapse rates, with significant severe wind having some of the steepest 0--3 and 2--4 km lapse rates in the dataset, leading to larger CAPE compared to cases without wind reports (not shown). When compared to severe wind cases, significant severe wind cases tend to have lower CAPE, however, these results are non-conclusive as the sample size for significant wind is only 12 cases. The freezing level heights are, however, significantly (\textit{p} $<$ 0.05) lower for significant severe wind reports compared to severe wind cases. Additionally, LCL height is slightly higher for cases with wind reports, with significant severe wind cases having higher LCLs than severe wind cases (not shown). The higher CAPE and LCLs likely lead to deeper cold pools from enhanced water loading and evaporational cooling below the cloud layer \citep{Srivastava_Evaporatively_1985, wakimoto_DryMicroburst_1985, markowski2010mesoscale}, and enhancing the downward transport of momentum. \par
\subsection{Skew-Ts}
    Excluding the subset of elevated cases as dissused in the methods section, the mean LM skew-T has 1974 J kg$^-$$^1$ of MUCAPE with moderately dry conditions from the surface to $\sim$900 hPa and deeper moisture roughly from 900 to 750 hPa (Fig. 6a). Low-level CAPE is not overly large (73.5 J kg$^-$$^1$ 0--3 km CAPE), with a higher concentration of CAPE from 700 hPa to the equilibrium level (EL; Figs. 5a and 6a,b). Similarly to the CAPE distribution plots (Figs. 3b,c), marginal LM cases have the weakest SB and MUCAPE, which increases with increasing mesoanticyclone strength (Figs. 6a,b). However, marginal MUCAPE and weak MUCAPE, and SBCAPE have the largest low-level CAPE up to $\sim$3.25 km above which, strong LM MUCAPE has the largest CAPE. Additionally, marginal LMs have the largest SBCIN, while MUCIN is generally consistent among mesoanticyclone intensities (Figs. 6e,f). These differences in the CAPE and CIN distributions are driven primarily by warmer overall profiles for increasing mesoanticyclone strength, and increases to dewpoint in both the low and mid-levels (Fig. 5b). As such, strong LMs tend to have deeper moisture throughout most of the profile (deepest through the lowest 500 hPa), whereas the marginal cases tend to be drier (driest through the lowest 500 hPa). This results in large differences in the LFC and EL heights and subsequent total profile CAPE. For example, strong cases, which tend to have the largest CAPE, have the lowest LFC and highest EL heights, whereas marginal LMs, which tend to have the lowest CAPE, have the highest LFC and lowest EL heights (Figs. 5b and 6a,b). \par
    \begin{figure}
    \centering
    \includegraphics[width=0.875\linewidth]{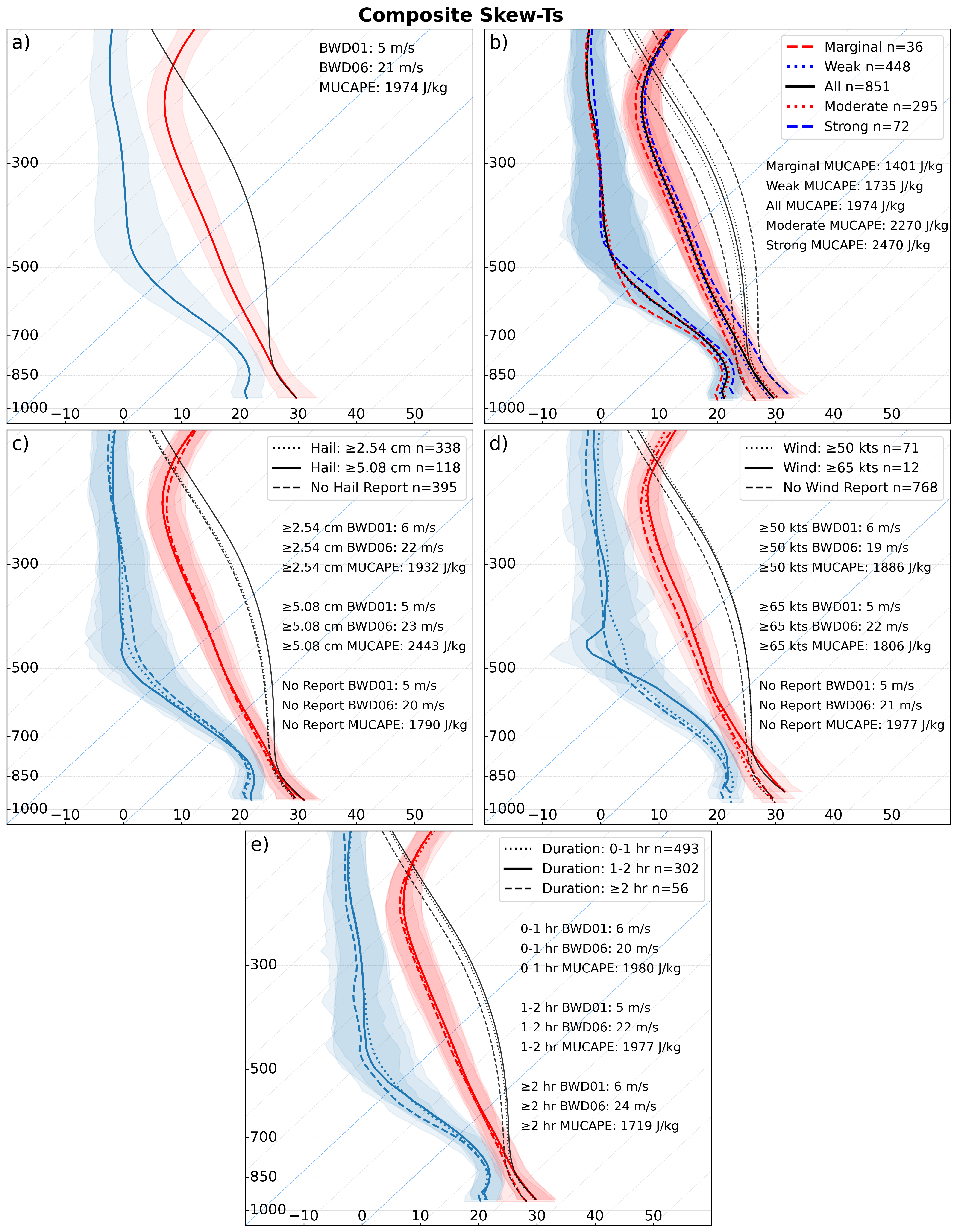}
    \caption{Composite skew-Ts for a) the mean of the 851 LM cases, b) each mesoanticyclone strength, c) cases with hail reports divided by severe, significant severe and cases without hail reports, d) cases with wind reports divided by severe, significant severe and cases without wind reports, and e) cases divided by duration lasting 0--1 hour, 1--2 hours, and $\ge$ 2 hours. Shaded is the area between the 10\textsuperscript{th} and 90\textsuperscript{th} percentiles for the temperature (red) and dewpoint (blue). For each subfigure, the median MUCAPE and bulk wind difference for 0--1 km (BWD01) and 0--6 km (BWD06) layers are included, except for b), which only has MUCAPE for each intensity.}
    \label{fig:5}
\end{figure}
\begin{figure}
    \centering
    \includegraphics[width=1\linewidth]{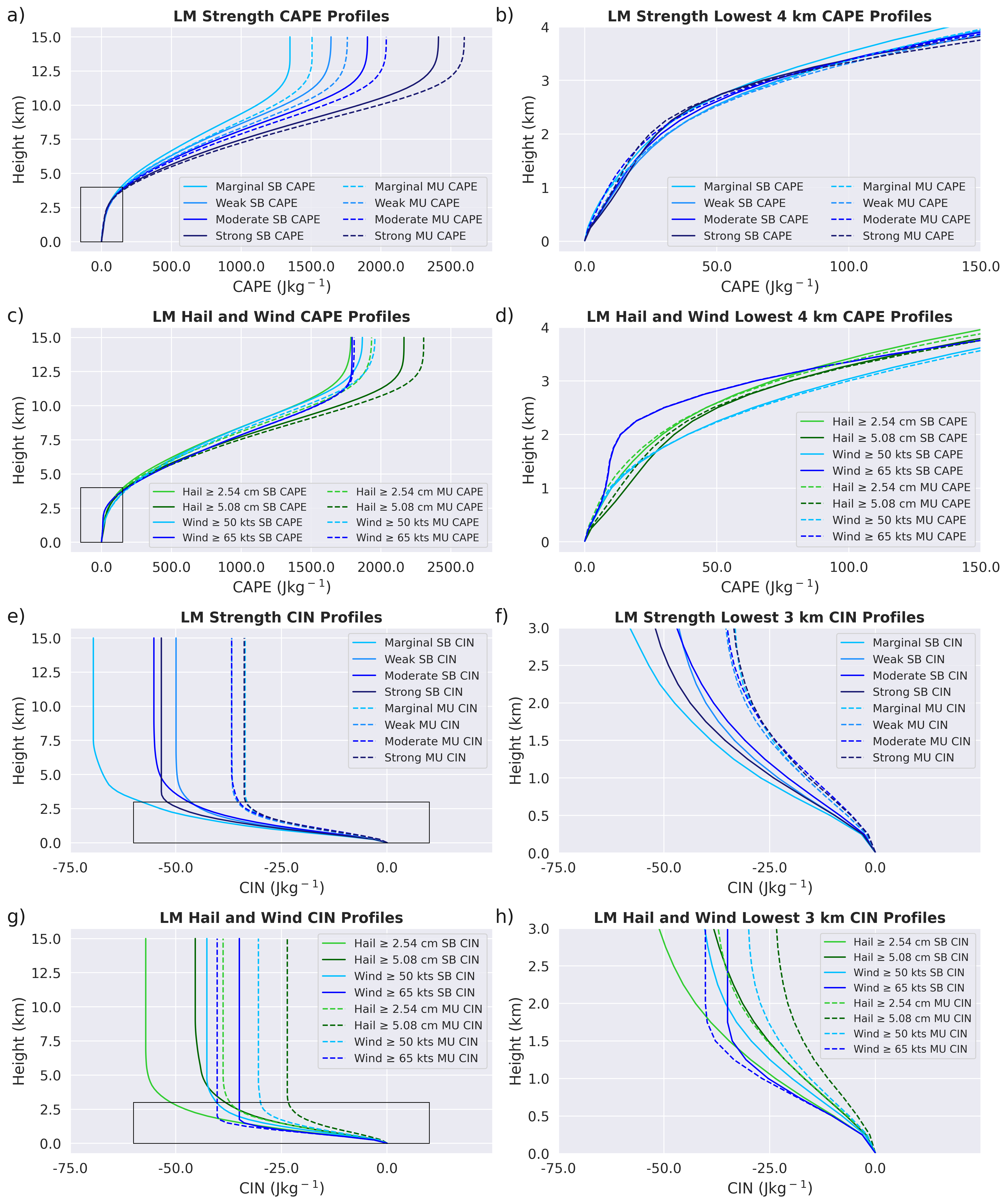}
    \caption{Mean SB and MUCAPE and CIN with respect to height separated by mesoanticyclone strength, hail size, and wind speed. The first column contains the full vertical profiles with rectangles that outline the corresponding zoomed in areas in the second column.}
    \label{fig:6}
\end{figure}
    A possible explanation for why marginal LMs may be weaker is from the dilution of the updraft decreasing the overall strength \citep[e.g.,][]{zipser_views_2003, romps_undiluted_2010, romps_nature_2010}. However, marginal LMs tend to have higher shear, leading to a smaller fraction of CAPE loss \citep[e.g.,][]{peters_role_2019, peters_are_2020}. Another potential explanation is that these cases tend to be consistent with the high-shear-low-CAPE (HSLC) archetype \citep{sherburn_2014, sherburn_2016, wade_dynamics_2021}, which have shallower overall storm depth and mesocyclones closer to the surface. However, RM HSLC cases often have larger CAPE concentrated within the lowest few kilometers \citep{sherburn_2014, sherburn_2016, wade_dynamics_2021}, which is not the case for these LM cases (65.2 J kg$^-$$^1$ 0--3 km CAPE). Most CAPE in marginal LMs tends to be concentrated between $\sim$5--10 km (Figs. 5b and 6a), potentially resulting in taller mesoanticyclones compared to RM HSLC cases, which suggests this is less likely to be a detection bias. Furthermore, cases in this dataset were required to remain within 200 km of a WSR-88D radar and within 150 km for at least 75\% of their lifespan, making it less likely that their mesoanticyclones were located below the radar beam height. Radar observations and numerical simulations are required to further investigate the internal dynamics of these marginal LMs and determine whether they are physically consistent with other HSLC cases.\par
    Skew-Ts for LMs that last 0--2 hours are generally similar, whereas those that exceed 2 hours have stronger shear and lower CAPE (Fig. 5e). These profiles have modestly cooler and slightly drier boundary layers, a factor that becomes more pronounced between 700--200 hPa in comparison to the other categories. For longer-lived LMs, the 0--6 km bulk wind difference increases while CAPE generally decreases. The LFCs for cases $\ge$2 hours are higher, and similar to the marginal cases, the drier mid-levels may act to dilute the updraft. However, higher shear may act against dilution due to a smaller fraction of CAPE loss \citep[e.g.,][]{peters_role_2019, peters_are_2020}. Additionally, these cases tend to have little low-level CAPE, with a larger distribution of this CAPE above the freezing level, potentially leading to a more favorable environment for hail growth \citep{nixon_hodographs_2023}.\par
    When comparing skew-Ts with hail reports, the mean skew-T for LMs with significant severe hail has the largest CAPE and aligns with the optimal range for hail growth \citep{lin_influences_2022}. These cases also have lower 0--1 km shear, deeper boundary layer moisture, drier mid-levels ($\sim$700--300 hPa), and higher ELs (Fig. 5c). Additionally, significant severe hail cases have the largest low-level CAPE and lowest CIN (both SB and MU), signifying that these cases are more likely to be surface-based (Fig. 6). Furthermore, these cases have higher average CAPE ({$>$} 145.3 J kg$^-$$^1$) concentrated below the hail growth region; however, this may be offset by the weaker low-level shear \citep{nixon_hodographs_2023}. Cases with severe hail have the weakest CAPE and highest shear with a drier boundary layer and deeper mid-level ($\sim$700--300 hPa) moisture (Fig. 5c). Moreover, these cases have the lowest low-level CAPE with the MUCAPE being the lowest of all categories below 1.5 km (Figs. 6c,d). Severe hail cases also have greater SBCIN and closely align with the strong LM cases, while their low-level MUCIN closely aligns with the SBCIN for significant severe hail cases (Figs. 6e,f,g,h). The cases without hail reports closely align with the severe hail cases, except they exhibit weaker shear and slightly greater mid-level moisture.\par
    The mean skew-T for significant severe wind has the largest dewpoint depression, including the driest surface-to-$\sim$875 hPa (Fig. 5d). These cases would tend to favor greater subcloud depth, with moisture predominantly found above the mixed layer through 550 hPa. There is a dry pocket from $\sim$500 hPa to 400 hPa within the composite and while this can aid in initiating and strengthening wind gusts at the surface, given the small sample size, this is likely a function of outliers as seen in driest percentile (Fig. 5d). Additionally, significant severe wind cases have the lowest 0--3 km CAPE (SB and MU) out of all cases with only $\sim$35 J kg$^-$$^1$ in the lowest 2.5 km (Figs. 6a,b,c,d). The surface elevation is highest for significant severe wind cases, likely a function of geographic prevalence—9 of the 12 cases are in the High Plains and one is in Arizona \citep{van_den_broeke_climo_LM_2025}. Storms in these regions typically have drier surface conditions, leading to higher bases, enhancing evaporational cooling and outflow. As for LMs with severe wind, their skew-Ts have the highest low-level moisture out of the three categories but are still generally dry from the surface to $\sim$900 hPa (Fig. 5d). The cases with severe wind have deeper boundary layer moisture above 900 hPa, closely aligning with the cases without wind reports until about 500 hPa, where they have the highest moisture out of the three categories until about 400 hPa. Generally, the mean parcel profile for the severe wind cases matches the significant severe wind parcel profile, except for lower expected cloud bases and higher low-level CAPE (Fig. 5d). Finally, the LM cases with no wind reports tend to have drier boundary layer conditions than severe wind cases; however, above the boundary layer, their profiles share similarities except for slightly drier and cooler throughout most of no wind reports profile (Fig. 5d).\par
\subsection{Kinematics}
    Unlike most parameters, 0--1 km shear is highest for marginal LM cases, decreasing with increasing mesoanticyclone strength (Fig. 7a). There is little difference in 0--6 km shear \citep[Fig. 7d;][]{Bunkers_2024_motion}. Instead, the effective bulk wind difference (EBWD), which accounts and corrects for elevated storms, may be a better deep layer shear metric for LMs than 0--6 km shear. As such, EBWD tends to be slightly higher with increasing mesoanticyclone strength, except for marginal cases, which tend to have the highest bulk shear (Fig. 7g). With increasing duration, EBWD has little variation among intensities; however, 0--6 km shear increases following previous literature \citep[not shown; e.g.,][]{bunkers_observational_2006-1, davenport_environmental_2021, Bunkers_2024_motion}. \par
\begin{figure}
    \centering
    \includegraphics[width=1\linewidth]{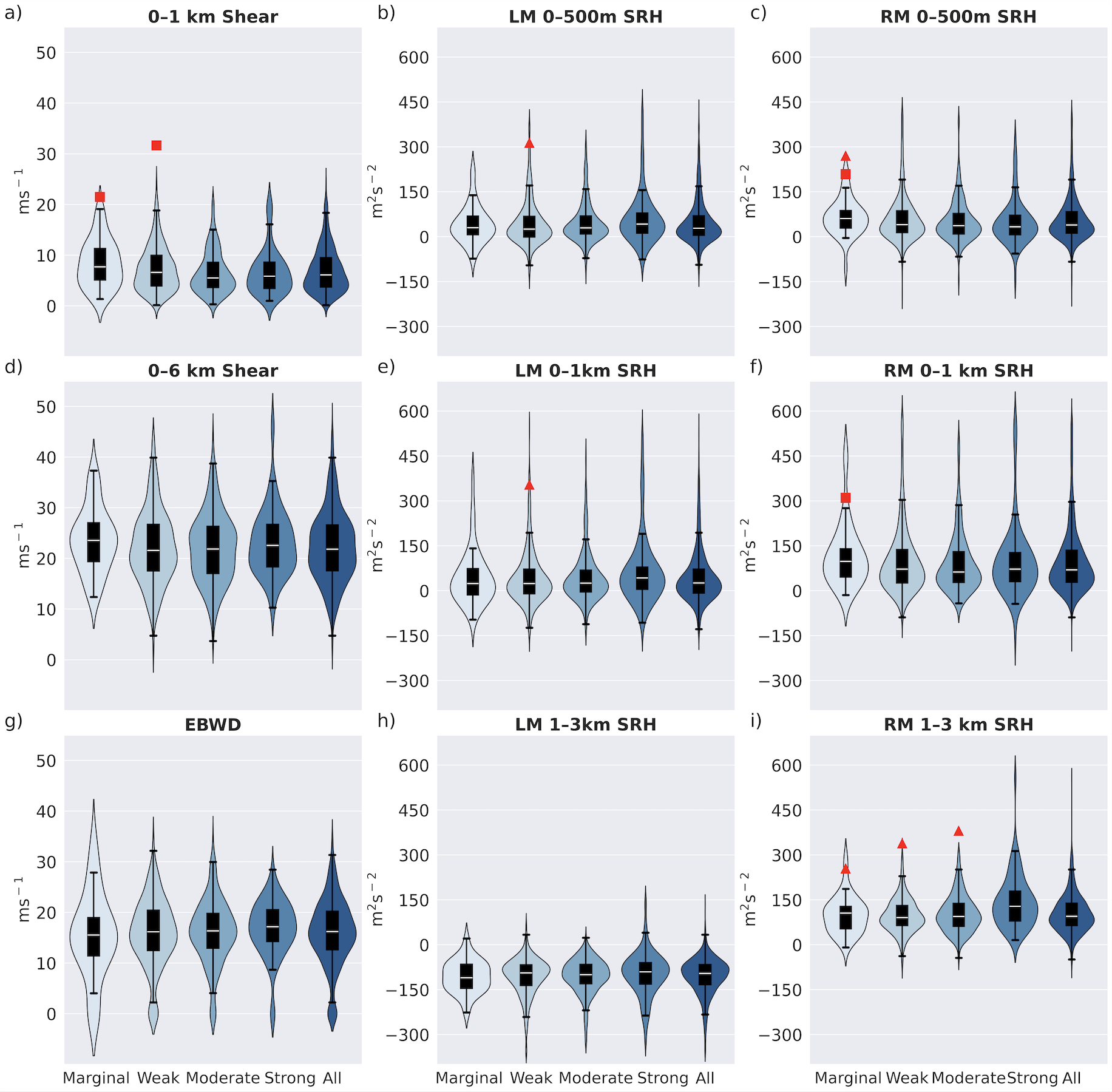}
    \caption{As for Fig. 3, except for kinematic variables.}
    \label{fig:7}
\end{figure}
   SRH quantifies streamwise vorticity and storm-relative inflow that support low- and mid-level mesocyclones, making it a valuable predictor of updraft rotation, strength, and tornadic potential \citep[e.g.,][]{davies-jones_1990_SRH, davies_and_johns_1993_wind_shear_and_helicity, esterheld_discriminating_2008, goldacker_assessing_2023}. For LMs, SRH (both LM and RM) 0--0.5, 0--1, 0--3, and 1--3 km show little variation as a function of mesoanticyclone intensity, but there is sensitivity between the different SRH layers [e.g., 0--1 compared to 0--3 km; Figs. 7b,c,e,f,h,i; not shown]. Marginal LMs tend to have more positive SRH, which may result in weaker updrafts due to the “negative” curvature for anticyclonic rotation. However, as previously discussed, marginal LMs tend to be elevated and therefore less likely to be affected by the lowest kilometer of positive SRH \citep{thompson_effective_2007}. Furthermore, this appears to apply to much of the sample, since most LMs tend to have unfavorable SRH in the lowest kilometer (Figs. 7b,c,e,f). This could have a couple of impacts: 1) there is a higher tendency for elevated LMs in most environments, and/or 2) LMs are weakened because of this low-level positive SRH. These are supported as RM SRH tends to be larger than LM SRH (Figs. 7b,c,e,f,h,i), signifying that LMs tend to form in low-level environments more supportive of RMs (i.e., cyclonic rotation), consistent with the prevalent climatological wind field in the CONUS \citep{homeyer_2025}. As such, commonly used surface-based SRH layers are not useful predictors for LM potential and intensity. Instead, 1--3 km SRH better differentiates between the two (Figs. 7h,i), and LM 1--3 km SRH predicts LM potential better than the other LM SRH layers \citep{bunkers_vertical_2002, edwards_assessment_2004, Bunkers_2024_motion}.\par
    Differentiating between severe hail and significant severe hail cases, the only significant (\textit{p} $<$ 0.05) differences are 0--1 km shear and LM 1--3 km SRH, which tend to be weaker for significant severe hail (Figs. 8a,d,g; not shown). The only noticeable difference when comparing significant hail cases to the rest of the dataset without hail reports is deep layer shear, which tends to be larger (not shown). LM cases with significant severe wind tend to have weaker 0--1 km shear, higher 0--6 km shear, and weaker SRH for all three layers (not shown).\par
\begin{figure}
    \centering
    \includegraphics[width=1\linewidth]{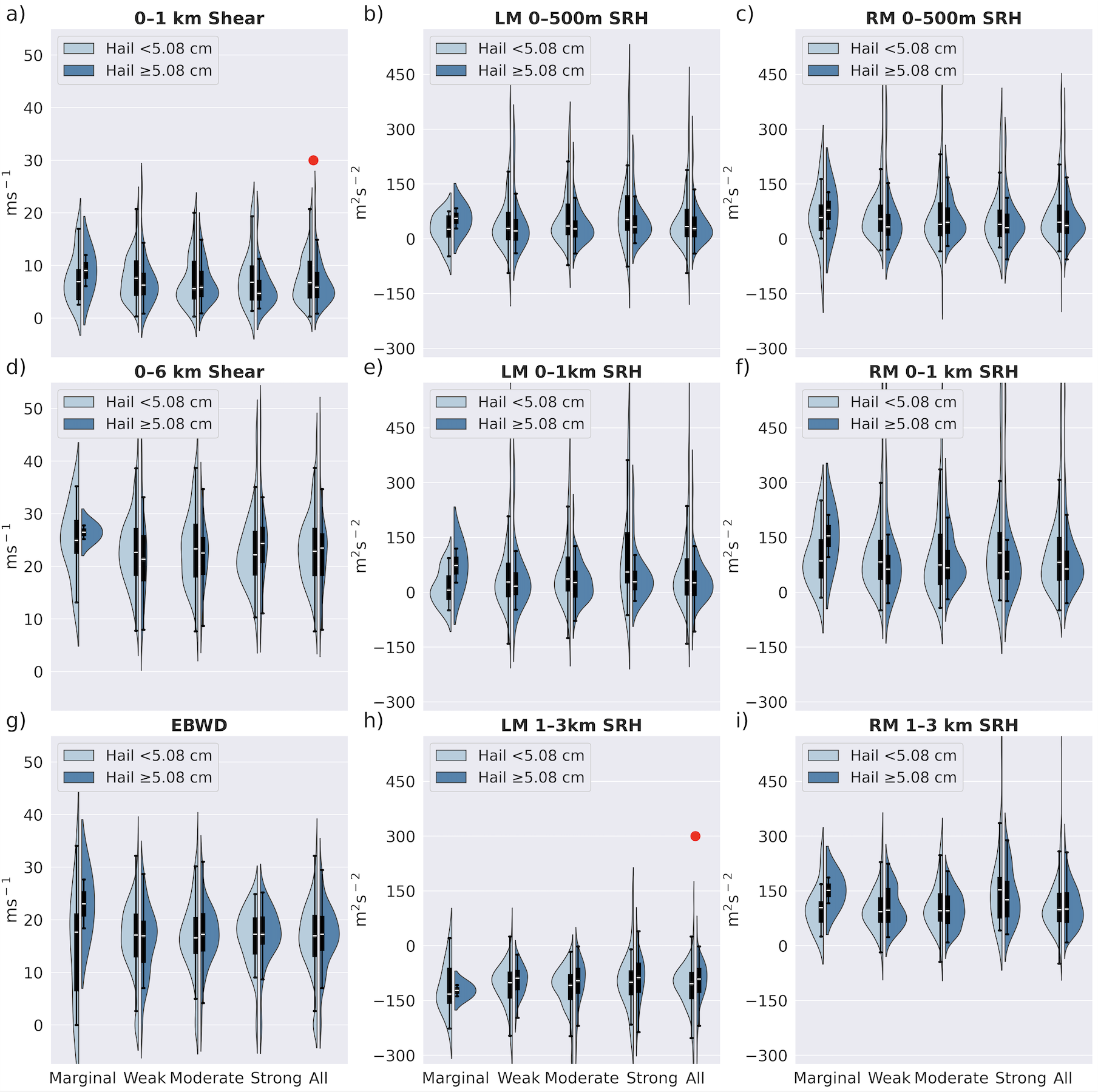}
    \caption{As for Fig. 5, except for several kinematic variables.}
    \label{fig:8}
\end{figure}
\subsection{Hodographs}
    Since commonly used shear parameters show little skill in predicting LM intensity, we explore whether the hodograph shape and size may be revealing. \citet{bunkers_vertical_2002}—one of the only studies to look at hodograph shape for LMs—divided each LM hodograph into two groups using the shear vector turning in the lowest 3 km. The first group were classified as clockwise dominated hodographs, accounting for 37\% of that dataset, and the second group were classified as straight hodographs, accounting for the remaining 63\%. Our results are very similar; 35.2\% of our cases have clockwise-curved hodographs, while the remaining 64.8\% are classified as straight. \par
    The mean hodographs for all mesoanticyclone strengths show small differences in overall shape (Fig. 9a). All classes follow the veer-back-veer or reverse S-shape hodograph found in prior studies \citep{bunkers_vertical_2002, bunkers_documentation_2007, bocheva_severe_2018, nixon_distinguishing_2022, Bunkers_2024_motion}. In the marginal hodograph, veering in the 9--12 km layer may influence the distribution of precipitation (Fig. 9a). Since most LMs move northeast \citep{van_den_broeke_climo_LM_2025}, this potential precipitation displacement may lead to a broader inflow region as the precipitation would be distributed towards the south and away from the storm motion. The low levels exhibit considerable veering and thus positive cyclonic SRH, as discussed above. While more than 60.0\% of the hodographs are classified as straight, they still exhibited sufficient clockwise curvature in the low levels to skew the mean low-level shape. This clockwise curvature results in positive streamwise vorticity, which is less supportive of anticyclonic rotation \citep{davies-jones_streamwise_1984}. \par
\begin{figure}
    \centering
    \includegraphics[width=0.925\linewidth]{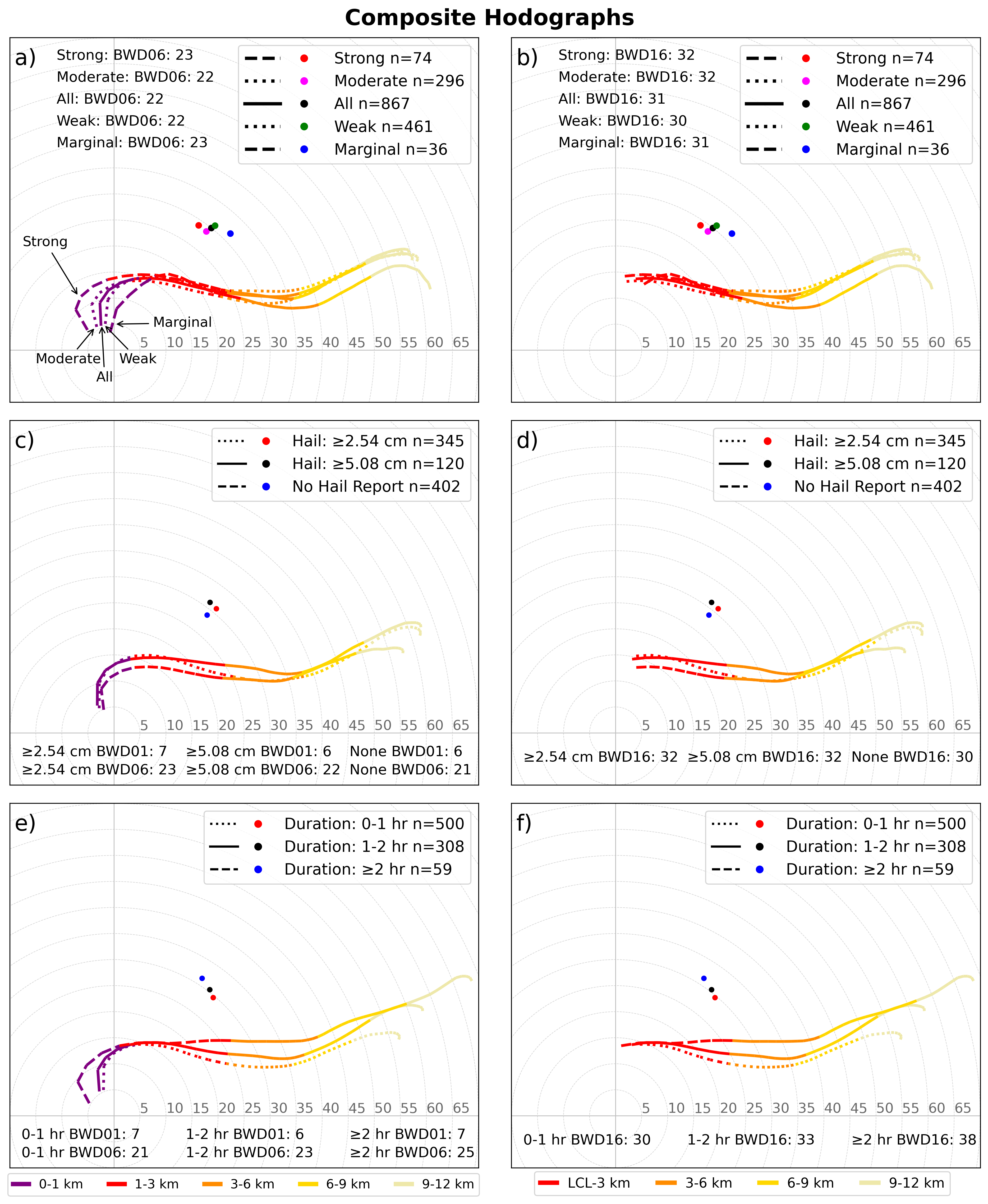}
    \caption{Composite hodographs for a) all mesoanticyclone strengths and mean dataset hodograph, c) cases with hail reports divided by severe, significant severe, and cases without hail reports, and e) cases divided by duration length including those lasting 0--1 hour, 1--2 hours, and those greater than or equal to 2 hours. For c) and e), the bulk wind difference for 0--1 km (BWD01) and 0--6 km (BWD06) layers are included, whereas a) only has the BWD06. b), d), and f) are the same as a), d), and e) respectively, except for the removal of data below the LCL height and adjusted BWD for the 1--6 km layer.}
    \label{fig:9}
\end{figure}
    Recall that with increasing mesoanticyclone strength, LMs have higher CAPE and LCLs, with strong cases having the highest median values (Figs. 3b,c,h). Cases with higher CAPE and LCLs likely have deeper cold pools driven by greater precipitation loading and extensive evaporational cooling below the cloud layer, with the potential implication being that more of the low-level hodograph is not being realized by the storm. This would result in less of the clockwise curvature being relevant to the inflow. Restricting the profile to the portion above the LCL heights, the hodograph is more consistent with expectations of anti-streamwise vorticity (Fig. 9b). Although there is still clockwise curvature at and slightly above the LCL height, especially for the marginal cases, the middle portion (e.g., 2--8 km) of the hodographs have counterclockwise turning with height (i.e., anti-streamwise vorticity; Fig. 7h). Since 50\% of the marginal cases have an MU level greater than one, these storms are likely elevated and may not realize the clockwise curvature at and above LCL height \citep{thompson_effective_2007}.\par
    Given the “better-looking” profile for LMs by removing data below the LCL height, LMs may tend to have inflow above and across their cold pools, and/or infrequently have lower-level mesoanticyclones due to the presence of weak to moderately positive SRH in the lower levels. Without the presence of a lower-level mesoanticyclone and subsequent induced vertical PGF, this may mean LMs are more transient and weaker than RMs in the same environment, especially HSLC cases that often depend on the low-level mesocyclone \citep{sherburn_2014, sherburn_2016, wade_dynamics_2021}. Furthermore, this may partly explain the rarity of tornado production in LMs [see review by \citet{fischer_supercell_2024}]. Most of the observed tornadic LM cases are of the landspout or hybrid variety rather than mesoanticyclonic or are closely associated with boundaries \citep{hammond_study_1967, brown_iowa_1980, monteverdi_first_2001, dostalek_aspects_2004, bunkers_documentation_2007}. However, this is not always the case, as there have been higher-end anticyclonic tornadoes likely resulting from LMs with a low-level mesoanticyclone \citep{brown_iowa_1980, monteverdi_first_2001}. This suggests that radar observations and/or modeling studies are required to confirm how frequently LMs have low-level mesoanticyclones.\par
    Separating hodographs by cases that produced severe hail, significant severe hail, and those that did not have any hail reports leads to little difference in overall hodograph structure (Fig. 9c). The largest difference is the amount of backing aloft, especially in the 6--9 km layer, where cases without hail reports have less backing and are straighter versus those with hail reports, which tend to have stronger backing aloft, especially for severe hail. Additionally, the veering in the 9--12 km layer is the most extreme for the cases with severe hail and least extreme for the cases without hail reports (Fig. 9c). Restricting the hodograph with the lowest level at LCL height, the overall structure, especially throughout the layer from LCL to 6 km, is straight, and the cases with hail reports tend to have larger 3--6 km shear (Fig. 9d). Straight hodographs lead to an elongation of the updraft width in the east-west direction and can lead to longer residence time for hail embryos, leading to better conditions for hail growth \citep{dennis_impact_2017, kumjian_hail_2020, kumjian_evolution_2021, pounds_analysis_2024}. Additionally, this elongation of the updraft may support longer-lived LMs as well, since longer-lived LMs have longer and straighter hodograph structures, leading to larger storm-relative inflow \citep[Figs. 9e,f; e.g.,][]{bunkers_observational_2006-1, davenport_environmental_2021}.\par
\subsection{Composite Parameters}
    While most parameters struggle with predicting LM strength, there is at least some skill for CAPE, LCL height, 1--3 SRH, and to a lesser extent 0--1 km shear, and EBWD. Combining several of these may provide a useful composite parameter to predict LMs as the current LM supercell composite parameter (SCP), which is derived from MUCAPE, effective-SRH, EBWD, and MUCIN, has no skill in predicting LMs and LM mesoanticyclone intensity (Fig. 10a). The RM SCP does a better job at predicting and differentiating between LM strength categories (Fig. 10b). The main reason for this is likely because LMs occur in environments that support RMs and there is little difference between the two in parameter space. The current SCP relies on the difference in the effective-SRH, and since LMs tend to have positive SRH in at least the lowest kilometer, the SCP does not differentiate well between the two types of supercells. What appears to better differentiate the two types of supercells is the hodograph shape and 1--3 km SRH. The current SCP does not take this into account and we intend to address this in future work, as this requires a more detailed analysis of the spatio-temporal variability of environments given the propensity for cases to be related to boundaries. For now, we advise against operational use of LM SCP based on the results presented here. \par
\begin{figure}
    \centering
    \includegraphics[width=1\linewidth]{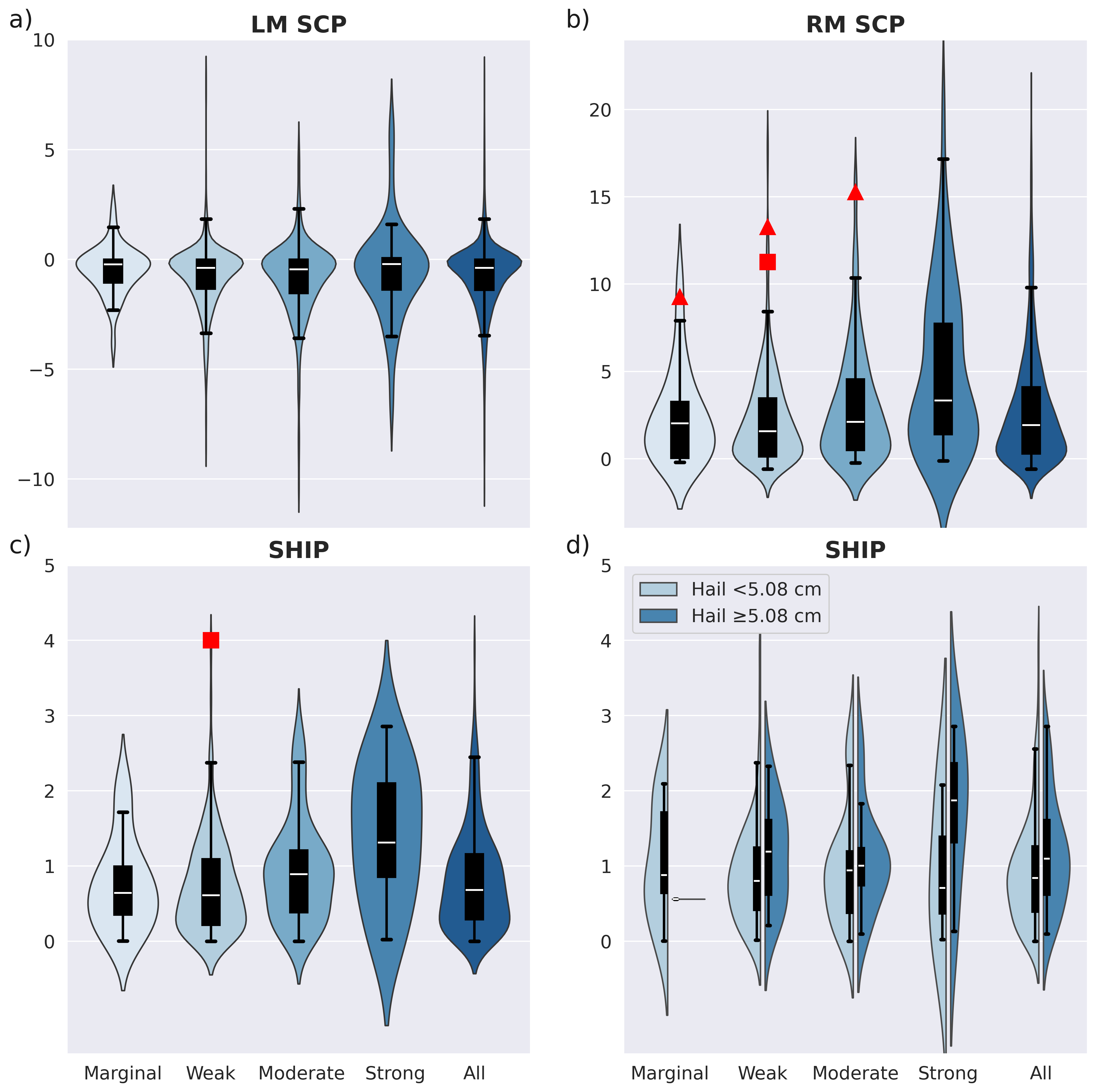}
    \caption{Composite parameters for the LM dataset. a) is the LM supercell composite parameter, b) is the RM supercell composite parameter, c) is the significant hail parameter, and d) is the significant hail parameter divided into two groups: severe hail and significant severe. Statistical significance testing and sample sizes are the same as for Figs. 3 and 5.}
    \label{fig:10}
\end{figure}
    Since LMs tend to predominantly produce hail, the significant hail parameter (SHIP) was calculated for each sounding. SHIP increases with mesoanticyclone intensity (Fig. 10c), following the distributions of MUCAPE and 700--500 hPa lapse rates, which are part of the formulation of SHIP. Breaking down all hail cases, the SHIP parameter does statistically differentiate between populations, particularly for the moderate and strong cases; however, these are not consistent enough to rely solely on this parameter to predict significant hail from LMs, consistent with its overall performance for hail \citep[Fig. 10d;][]{gensini_hail_2021}. \par

\section{Conclusions}
    This study uses a dataset of 889 observed and quality-controlled LMs, spanning across North America from 2011--2022 \citep{van_den_broeke_left-moving_2024}. Each case has assigned mesoanticyclone strength (marginal, weak, moderate, and strong) based on the maximum anticyclonic rotation following the \citet{andra_1997} nomogram. RAP/RUC sounding profiles were obtained for 867 of these cases (22 cases had missing profiles) to compute the nearest inflow parameter space. Commonly used convective metrics, such as lapse rates, CAPE, CIN, LCL height, bulk shear, and SRH, in addition to many others, were calculated for these available profiles in addition to hodograph and skew-T construction to establish a baseline climatology of LM environments. All metrics were broken down by mesoanticyclone strength, hail size, wind speed, and duration of the event to explore potential sources of predictability for these events. \par
    Results suggest that LM environments tend to overlap with RM environments. As such, LMs form in a wide range of CAPE values with the ML interquartile range from $\sim$750--2000 J kg$^-$$^1$ and generally highest with increasing mesoanticyclone strength. LM bulk shear metrics are similar to those of RMs and have few differences between LM mesoanticyclone strengths. LM and RM SRH tend to be positive in the lowest kilometer, with varying SRH layers (0--0.5, 0--1, 0--3, and 1--3 km) having little skill in predicting LM strength and potential. The LM 1--3 km SRH best represents LM potential and should be used instead of effective-SRH to forecast LMs. The mean LM hodograph matches prior research with a veer-back-veer or reverse S-shaped hodograph. \par
    Given the lack of anti-streamwise vorticity in the lowest kilometer, we hypothesize that LMs 1) are weakened by the positive SRH, 2) have infrequent or weak lower-level mesoanticyclones, and 3) tend to have inflow above the lowest kilometer (i.e., at LCL height and above). These hypotheses are supported when restricting the hodographs to the LCL height. The hodographs become more favorable for anticyclonically rotating updrafts given the straighter nature (i.e., crosswise vorticity) and backing in the middle portions (i.e., anti-streamwise vorticity). Furthermore, the veering and lack of anti-streamwise vorticity in the low levels most likely retards low-level mesoanticyclone formation, decreasing the potential for tornadoes. Based on case literature, anticyclonic tornadoes can form from LMs, but are more likely formed along the gust front, boundaries, or from the cyclonic vortex pair underneath the updraft. \par
    Since the current SCP solely relies on the sign of the effective SRH to differentiate between LMs and RMs, the LM SCP has no skill in predicting LMs, whereas the RM SCP is better at predicting LM strength. Therefore, we advise not to use the current LM SCP to predict LMs and instead base prediction on the LM 1--3 km SRH and the hodograph structure, especially that above the LCL height. Moreover, the RM SCP may provide utility for forecasting LM strength, but may not be the optimal parameter to do so. New parameters may be needed to assess LM potential.\par
   Since LMs tend to move poleward in the Northern Hemisphere, they may start surface-based and become elevated after crossing towards the cool side of a boundary. With over 40\% of LMs with a most-unstable level above the surface in the nearest inflow sounding, we expect the number of cases that become elevated to increase if effectively sampled at a later point in their lifetimes. Successfully transitioning to an elevated state on the cool side of a boundary may aid in the sustenance of a LM due to the increasingly favorable shear profile for anticyclonic rotation \citep[e.g.,][]{bunkers_observational_2006-1}. This is supported as more than 50\% of LMs with analysis periods $\ge$ 2 hours are elevated, and likely lived longer as they move into a more anticyclonically favorable environment (i.e., the loss of low-level positive SRH). Further research is needed to explore if LM proximity to and crossing of boundaries may be useful in predicting LM potential, severity, and longevity. In RMs, this transition to a more stable environment can either be helpful or detrimental to the storm \citep{nowotarski_characteristics_2011, coffer_ealry_evening_transition_2015, macintosh_6_2017, gropp_NocturnalTransition_2018, brown_EarlyEveningTransition_2021}, but appears to impact the strength of the low-level mesocyclone. LMs may have a smoother transition across a stable boundary as they are less reliant on this feature and have pre-existing inflow regions that are above the surface. However, \citet{grasso_dissipation_2000} suggests that LMs may dissipate quickly after splitting because the storms' inflow region is dominated by the downdraft-cooled and negatively buoyant air. While this may be the case for some LMs, especially shorter-lived events, further numerical simulations are needed to test our hypotheses and understand the dynamics of LMs, how they physically differ from RMs, and why they predominantly produce hail and rarely produce tornadoes.\par
    The favorable environment for large hail growth, especially if the lowest kilometer is not realized by the storm, may be one reason why LMs predominantly produce hail. LMs with hail reports tend to have steeper lapse rates, larger CAPE, higher LCLs, higher freezing levels, slightly higher 0--6 km shear and EBWD, slightly weaker 0--1 km shear, and weaker LM 1--3 km SRH. The biggest difference in the hodograph structure is that storms with larger hail tend to have stronger shear in the east-west direction, especially in the 3--6 km layer. This is likely to elongate the updraft west to east, allowing for longer residence time in the hail growth region \citep{dennis_impact_2017, kumjian_hail_2020, kumjian_evolution_2021, nixon_hodographs_2023}. Furthermore, there is larger CAPE (largest for significant severe hail cases) within the hail growth region, which has been shown in large hail-producing RMs \citep{nixon_hodographs_2023}, potentially leading to rapid hail growth. Lastly, elevated LMs tend to also have favorable profiles for hail generation and account for over 40\% of all hail cases, with weaker CAPE concentrated mainly in the hail growth region, and stronger bulk shear. \par
    A potential bias within this work is that the results are based on the single point, nearest inflow RUC/RAP sounding profile for each case. We acknowledge the biases that arise from using a single model and single point sounding from a subjective analysis of the storm’s centroid and inflow region. An analysis of the profiles through each storm's life may be useful in determining the environmental evolution. As discussed above, the proximity to boundaries is of particular interest given that a transition to elevated storms may aid in LM duration. Additionally, the analysis for the hail and wind reports are based on the nearest inflow sounding at the beginning of the storm's classification and not the closest sounding to the report. A more representative profile may be the nearest soundings for all of the reports (though these would still not reflect storm modification), which future work could address. We intend to compare the results presented here against other profile data to address whether the changing nature of RAP/RUC (e.g., changing vertical levels and horizontal resolution) through time influences the robustness of the associations shown here.\par
    Lastly, a similar comprehensive climatology of RMs with quantified mesocyclone strength has not yet been developed and would extend the knowledge on the parameter space of RMs and could be compared directly to the research presented here and in \citet{van_den_broeke_climo_LM_2025}. Furthermore, comparing environments (e.g., synoptic, mesoscale, and parameter space) that support strong LMs and weak RMs, weak LMs and strong RMs, equally both components, and similarly for splitting and non-splitting events would help increase forecaster awareness and predictability of the associated hazards. \par

\clearpage
\acknowledgments
We thank Cynthia Van Den Broeke, Simon Kirby, Raychel Nelson, Benjamin Schweigert, Amy Conner, and Ashton Cook for developing the manually compiled LM supercell dataset. We also thank Ryan Jewell and Rich Thompson for their help in gathering and developing the storm reports and environmental dataset. Lastly, we thank Matt Bunkers and Cynthia Van Den Broeke for valuable comments on the initial form of this manuscript. This work was supported by the Earth and Ecosystem Science PhD program at Central Michigan University and the National Science Foundation under Grant AGS-2218623. 

%
%
\datastatement
The LM dataset used in this study can be found at \url{https://zenodo.org/doi/10.5281/zenodo.10551964}, the storm reports can be accessed through the Storm Events Database (\url{https://www.ncdc.noaa.gov/stormevents/}), the environmental RUC/RAP data can be accessed from NCEI (\url{https://www.ncei.noaa.gov/products/weather-climate-models/rapid-refresh-update}), and the compiled RAP/RUC environmental data for the LM cases can be found at \url{https://doi.org/10.5281/zenodo.15939413}.

\bibliographystyle{ametsocV6}
\bibliography{references}

@STRING{AGU 	= "Amer.\ Geophys.\ Union"}

@STRING{AN        = "Astrophys.\ Norv."}

@STRING{MA        = "Meteor.\ Appl."}

@misc{MunichRe2025,
  author = {MunichRe},
  title = {Natural Disaster Figures 2024},
  year = {2025}, 
  howpublished = {\url{https://www.munichre.com/content/dam/munichre/mrwebsitespressreleases/MunichRe-Mediarelease2025-NatCat2024.pdf}},
  note = {Accessed 30-04-2025},
}

@misc{Aon2025,
  author = {Aon},
  title = {2025 Climate Catastrophe Insight},
  year = {2025},
  howpublished = {\url{https://assets.aon.com/-/media/files/aon/reports/2025/2025-climate-catastrophe-insight.pdf}},
  note = {Accessed 30-04-2025},
}

@article {gensini_hail_2021,
      author = "Vittorio A. Gensini and Cody Converse and Walker S. Ashley and Mateusz Taszarek",
      title = "Machine Learning Classification of Significant Tornadoes and Hail in the {United States} Using {ERA5} Proximity Soundings",
      journal = "Weather and Forecasting",
      year = "2021",
      publisher = "American Meteorological Society",
      address = "Boston MA, USA",
      volume = "36",
      number = "6",
      doi = "10.1175/WAF-D-21-0056.1",
      pages= "2143--2160",
      extra = "https://journals.ametsoc.org/view/journals/wefo/36/6/WAF-D-21-0056.1.xml"
}

@inproceedings{Hales1988,
  author    = {Hales, J. E., Jr.},
  title     = {Improving the watch/warning program through use of significant event data},
  booktitle = {Preprints, 15th Conf. on Severe Local Storms},
  year      = {1988},
  address   = {Baltimore, MD},
  publisher = {Amer. Meteor. Soc.},
  pages     = {165--168}
}

@article{homeyer_2025,
      author = "Cameron R. Homeyer and Matthew J. Bunkers and John T. Allen and Amanda M. Murphy",
      title = "United States Supercell Storms and Their Severity: A 14-yr Radar-Based Climatology",
      journal = "Journal of Applied Meteorology and Climatology",
      year = "2025",
      publisher = "American Meteorological Society",
      address = "Boston MA, USA",
      doi = "10.1175/JAMC-D-24-0185.1",
      url = "https://journals.ametsoc.org/view/journals/apme/aop/JAMC-D-24-0185.1/JAMC-D-24-0185.1.xml"
}

@article{van_den_broeke_climo_LM_2025,
	author = {{Van Den Broeke}, Matthew and {Van Den Broeke}, Cynthia and Schweigert, Benjamin},
    title = {Spatiotemporal Characteristics of Anticyclonic Supercells in the Contiguous {United States}},
    journal = {J. Appl. Meteor. Climatol., in review},
    year = {2025}
}

@article{brooks_climatology_sounding_2002,
author = {Craven, J.P. and Brooks, Harold},
year = {2004},
month = {01},
pages = {13--24},
title = {Baseline climatology of sounding derived parameters associated with deep, moist convection},
volume = {28},
journal = {Natl. Wea. Dig.}
}

@article {Rasmussen_climatology_sounding_parameters_1998,
      author = "Erik N.  Rasmussen and David O.  Blanchard",
      title = "A Baseline Climatology of Sounding-Derived Supercell and Tornado Forecast Parameters",
      journal = "Weather and Forecasting",
      year = "1998",
      publisher = "American Meteorological Society",
      address = "Boston MA, USA",
      volume = "13",
      number = "4",
      doi = "10.1175/1520-0434(1998)013<1148:ABCOSD>2.0.CO;2",
      pages= "1148--1164",
      extra = "https://journals.ametsoc.org/view/journals/wefo/13/4/1520-0434_1998_013_1148_abcosd_2_0_co_2.xml"
}

@article {Srivastava_Evaporatively_1985,
      author = "R. C.  Srivastava",
      title = "A Simple Model of Evaporatively Driven Downdraft: Application to Microburst Downdraft",
      journal = "Journal of Atmospheric Sciences",
      year = "1985",
      publisher = "American Meteorological Society",
      address = "Boston MA, USA",
      volume = "42",
      number = "10",
      doi = "10.1175/1520-0469(1985)042<1004:ASMOED>2.0.CO;2",
      pages= "1004--1023",
      extra = "https://journals.ametsoc.org/view/journals/atsc/42/10/1520-0469_1985_042_1004_asmoed_2_0_co_2.xml"
}

@article {wakimoto_DryMicroburst_1985,
      author = "Roger M.  Wakimoto",
      title = "Forecasting Dry Microburst Activity over the {High Plains}",
      journal = "Monthly Weather Review",
      year = "1985",
      publisher = "American Meteorological Society",
      address = "Boston MA, USA",
      volume = "113",
      number = "7",
      doi = "10.1175/1520-0493(1985)113<1131:FDMAOT>2.0.CO;2",
      pages= "1131--1143",
      extra = "https://journals.ametsoc.org/view/journals/mwre/113/7/1520-0493_1985_113_1131_fdmaot_2_0_co_2.xml"
}

@book{markowski2010mesoscale,
  title={Mesoscale meteorology in midlatitudes},
  author={Markowski, Paul and Richardson, Yvette},
  year={2010},
  pages = {292--300},
  publisher={John Wiley \& Sons}
}

@article{blair_radar-based_nodate,
	title = {A Radar-Based Assessment of the Detectability of Giant Hail},
	pages = {1--30},
    year ={2011},
    journal = {E-Journal of Severe Storms Meteorology},
	author = {Blair, Scott F and Leighton, Jared W and Barjenbruch, Brian L and Gargan, William P and Deroche, Derek R and Boustead, Joshua M},
	langid = {english},
}

@article{klemp_simulation_1978,
	title = {The Simulation of Three-Dimensional Convective Storm Dynamics},
	volume = {35},
	issn = {0022-4928, 1520-0469},
	extra = {https://journals.ametsoc.org/view/journals/atsc/35/6/1520-0469_1978_035_1070_tsotdc_2_0_co_2.xml},
	doi = {10.1175/1520-0469(1978)035<1070:TSOTDC>2.0.CO;2},

	pages = {1070--1096},
	number = {6},
	journal = {Journal of the Atmospheric Sciences},
	author = {Klemp, Joseph B. and Wilhelmson, Robert B.},
	extradate = {2023-07-09},
    year = {1978},
	date = {1978-06-01},
	note = {Publisher: American Meteorological Society
Section: Journal of the Atmospheric Sciences},
}

@article{rotunno_rotation_1985,
	title = {On the Rotation and Propagation of Simulated Supercell Thunderstorms},
	volume = {42},
	issn = {0022-4928, 1520-0469},
	extra = {https://journals.ametsoc.org/view/journals/atsc/42/3/1520-0469_1985_042_0271_otrapo_2_0_co_2.xml},
	doi = {10.1175/1520-0469(1985)042<0271:OTRAPO>2.0.CO;2},
	pages = {271--292},
	number = {3},
	journal = {Journal of the Atmospheric Sciences},
	author = {Rotunno, Richard and Klemp, Joseph},
	extradate = {2023-07-06},
    year = {1985},
	date = {1985-02-01},
	note = {Publisher: American Meteorological Society
Section: Journal of the Atmospheric Sciences},

}

@incollection{weisman_characteristics_1986,
	location = {Boston, {MA}},
	title = {Characteristics of Isolated Convective Storms},
	isbn = {978-1-935704-20-1},
	extra = {https://doi.org/10.1007/978-1-935704-20-1_15},

	pages = {331--358},
	booktitle = {Mesoscale Meteorology and Forecasting},
	publisher = {American Meteorological Society},
	author = {Weisman, Morris L. and Klemp, Joseph B.},
	editor = {Ray, Peter S.},
	extradate = {2023-07-06},
	year ={1986},
	langid = {english},
	doi = {10.1007/978-1-935704-20-1_15},
}

@article{peters_are_2020,
	title = {Are Supercells Resistant to Entrainment because of Their Rotation?},
	volume = {77},
	issn = {0022-4928, 1520-0469},
	extra = {https://journals.ametsoc.org/view/journals/atsc/77/4/jas-d-19-0316.1.xml},
	doi = {10.1175/JAS-D-19-0316.1},
	pages = {1475--1495},
	number = {4},
	journal = {Journal of the Atmospheric Sciences},
	author = {Peters, John M. and Nowotarski, Christopher J. and Mullendore, Gretchen L.},
	extradate = {2023-09-25},
	date = {2020-04-01},
    year = {2020},
	note = {Publisher: American Meteorological Society
Section: Journal of the Atmospheric Sciences},

}

@article{peters_role_2019,
	title = {The Role of Vertical Wind Shear in Modulating Maximum Supercell Updraft Velocities},
	volume = {76},
	issn = {0022-4928, 1520-0469},
	extra = {https://journals.ametsoc.org/view/journals/atsc/76/10/jas-d-19-0096.1.xml},
	doi = {10.1175/JAS-D-19-0096.1},
	pages = {3169--3189},
	number = {10},
	journal = {Journal of the Atmospheric Sciences},
	author = {Peters, John M. and Nowotarski, Christopher J. and Morrison, Hugh},
	extradate = {2023-09-25},
	date = {2019-10-01},
    year = {2019},
	note = {Publisher: American Meteorological Society
Section: Journal of the Atmospheric Sciences},

}

@inproceedings{davies-jones_1990_SRH,
    author = {Robert Davies-Jones and Donald W. Burgess and M Foster},
    title = {Test of helicity as a tornado forecast parameter},
    booktitle = {Preprints, 16th Conf. on Severe Local Storms, Kananaskis Park, AB, Canada, Amer. Meteor. Soc.},
    year = {1990},
    pages = {588--592}
}

@article{esterheld_discriminating_2008,
	title = {Discriminating between Tornadic and Non-Tornadic Supercells: A New Hodograph Technique},
	volume = {3},
	rights = {Copyright (c) 2008},
	issn = {1559-5404},
	extra = {https://ejssm.com/ojs/index.php/site/article/view/15},
	doi = {10.55599/ejssm.v3i2.15},
	shorttitle = {Discriminating between Tornadic and Non-Tornadic Supercells},
	pages = {1--50},
	number = {2},
	journal = {E-Journal of Severe Storms Meteorology},
	author = {Esterheld, John M. and Giuliano, Donald J.},
	extradate = {2025-03-08},
	date = {2008-07-08},
    year = {2008},
	langid = {english},
	note = {Number: 2},

}

@inbook{davies_and_johns_1993_wind_shear_and_helicity,
author = {Davies, Jonathan M. and Johns, Robert H.},
publisher = {American Geophysical Union (AGU)},
isbn = {9781118664148},
title = {Some Wind and Instability Parameters Associated With Strong And Violent Tornadoes: 1. Wind Shear And Helicity},
booktitle = {The Tornado: Its Structure, Dynamics, Prediction, and Hazards},
chapter = {},
pages = {573--582},
doi = {https://doi.org/10.1029/GM079p0573},
year = {1993}
}

@article{coniglio_verification_2012,
	title = {Verification of {RUC} 0–1-h Forecasts and {SPC} Mesoscale Analyses Using {VORTEX}2 Soundings},
	volume = {27},
	issn = {1520-0434, 0882-8156},
	extra = {https://journals.ametsoc.org/view/journals/wefo/27/3/waf-d-11-00096_1.xml},
	doi = {10.1175/WAF-D-11-00096.1},
	pages = {667--683},
	number = {3},
	journal = {Weather and Forecasting},
	author = {Coniglio, Michael C.},
	extradate = {2023-11-10},
	date = {2012-06-01},
    year = {2012},
	note = {Publisher: American Meteorological Society
Section: Weather and Forecasting},

}

@article{nixon_hodographs_2023,
	title = {Hodographs and Skew {Ts} of Hail-Producing Storms},
	volume = {38},
	issn = {1520-0434, 0882-8156},
	extra = {https://journals.ametsoc.org/view/journals/wefo/38/11/WAF-D-23-0031.1.xml},
	doi = {10.1175/WAF-D-23-0031.1},
	pages = {2217--2236},
	number = {11},
	journal = {Weather and Forecasting},
	author = {Nixon, Cameron J. and Allen, John T. and Taszarek, Mateusz},
	extradate = {2023-11-10},
	date = {2023-11-01},
    year = {2023},
	note = {Publisher: American Meteorological Society
Section: Weather and Forecasting},

}

@article{thompson_close_2003,
	title = {Close Proximity Soundings within Supercell Environments Obtained from the {Rapid Update Cycle}},
	volume = {18},
	issn = {1520-0434, 0882-8156},
	extra = {https://journals.ametsoc.org/view/journals/wefo/18/6/1520-0434_2003_018_1243_cpswse_2_0_co_2.xml},
	doi = {10.1175/1520-0434(2003)018<1243:CPSWSE>2.0.CO;2},
	pages = {1243--1261},
	number = {6},
	journal = {Weather and Forecasting},
	author = {Thompson, Richard L. and Edwards, Roger and Hart, John A. and Elmore, Kimberly L. and Markowski, Paul},
	extradate = {2023-11-27},
	date = {2003-12-01},
    year = {2003},
	note = {Publisher: American Meteorological Society
Section: Weather and Forecasting},

}

@article{benjamin_north_2016,
	title = {A {North American} Hoextray Assimilation and Model Forecast Cycle: The {Rapid Refresh}},
	volume = {144},
	issn = {1520-0493, 0027-0644},
	extra = {https://journals.ametsoc.org/view/journals/mwre/144/4/mwr-d-15-0242.1.xml},
	doi = {10.1175/MWR-D-15-0242.1},
	shorttitle = {A North American Hoextray Assimilation and Model Forecast Cycle},
	pages = {1669--1694},
	number = {4},
	journal = {Monthly Weather Review},
	author = {Benjamin, Stanley G. and Weygandt, Stephen S. and Brown, John M. and Hu, Ming and Alexander, Curtis R. and Smirnova, Tatiana G. and Olson, Joseph B. and James, Eric P. and Dowell, David C. and Grell, Georg A. and Lin, Haidao and Peckham, Steven E. and Smith, Tracy Lorraine and Moninger, William R. and Kenyon, Jaymes S. and Manikin, Geoffrey S.},
	extradate = {2023-11-30},
	date = {2016-04-01},
    year = {2016},
	note = {Publisher: American Meteorological Society 
            Section: Monthly Weather Review},
}

@article{nixon_distinguishing_2022,
	title = {Distinguishing between Hodographs of Severe Hail and Tornadoes},
	volume = {37},
	issn = {1520-0434, 0882-8156},
	extra = {https://journals.ametsoc.org/view/journals/wefo/37/10/WAF-D-21-0136.1.xml},
	doi = {10.1175/WAF-D-21-0136.1},

	pages = {1761--1782},
	number = {10},
	journal = {Weather and Forecasting},
	author = {Nixon, Cameron J. and Allen, John T.},
	extradate = {2024-03-19},
	date = {2022-09-26},
    year = {2022},
	note = {Publisher: American Meteorological Society
Section: Weather and Forecasting},

}

@article{jewell_evaluation_2009,
	title = {Evaluation of {Alberta} Hail Growth Model Using Severe Hail Proximity Soundings from the {United States}},
	volume = {24},
	issn = {1520-0434, 0882-8156},
	extra = {https://journals.ametsoc.org/view/journals/wefo/24/6/2009waf2222230_1.xml},
	doi = {10.1175/2009WAF2222230.1},
	pages = {1592--1609},
	number = {6},
	journal = {Weather and Forecasting},
	author = {Jewell, Ryan and Brimelow, Julian},
	extradate = {2024-03-19},
	date = {2009-12-01},
    year = {2009},
	note = {Publisher: American Meteorological Society
Section: Weather and Forecasting},

}

@article{kumjian_hail_2020,
	title = {A Hail Growth Trajectory Model for Exploring the Environmental Controls on Hail Size: Model Physics and Idealized Tests},
	volume = {77},
	issn = {0022-4928, 1520-0469},
	extra = {https://journals.ametsoc.org/view/journals/atsc/77/8/jasD200016.xml},
	doi = {10.1175/JAS-D-20-0016.1},
	shorttitle = {A Hail Growth Trajectory Model for Exploring the Environmental Controls on Hail Size},
	pages = {2765--2791},
	number = {8},
	journal = {Journal of the Atmospheric Sciences},
	author = {Kumjian, Matthew R. and Lombardo, Kelly},
	extradate = {2024-03-19},
	date = {2020-07-22},
    year = {2020},
	note = {Publisher: American Meteorological Society
Section: Journal of the Atmospheric Sciences},

}

@article{kumjian_evolution_2021,
	title = {The Evolution of Hail Production in Simulated Supercell Storms},
	volume = {78},
	issn = {0022-4928, 1520-0469},
	extra = {https://journals.ametsoc.org/view/journals/atsc/78/11/JAS-D-21-0034.1.xml},
	doi = {10.1175/JAS-D-21-0034.1},
	pages = {3417--3440},
	number = {11},
	journal = {Journal of the Atmospheric Sciences},
	author = {Kumjian, Matthew R. and Lombardo, Kelly and Loeffler, Scott},
	extradate = {2024-03-19},
	date = {2021-11-01},
    year = {2021},
	note = {Publisher: American Meteorological Society
Section: Journal of the Atmospheric Sciences},

}

@article{jones_short-term_2017,
	title = {Short-term Forecasts of Left-Moving Supercells from an Experimental {Warn}-on-{Forecast} System},
	volume = {05},
	issn = {23256184},

	doi = {10.15191/nwajom.2017.0513},
	pages = {161--170},
	number = {13},
	journal = {Journal of Operational Meteorology},
	shortjournal = {J. Operational Meteor.},
	author = {Jones, Thomas A. and Nixon, Cameron},
	extradate = {2024-03-20},
	date = {2017-09-01},
    year = {2017},

}

@article{davies-jones_streamwise_1984,
	title = {Streamwise Vorticity: The Origin of Updraft Rotation in Supercell Storms},
	volume = {41},
	issn = {0022-4928, 1520-0469},
	extra = {https://journals.ametsoc.org/view/journals/atsc/41/20/1520-0469_1984_041_2991_svtoou_2_0_co_2.xml},
	doi = {10.1175/1520-0469(1984)041<2991:SVTOOU>2.0.CO;2},
	shorttitle = {Streamwise Vorticity},
	pages = {2991--3006},
	number = {20},
	journal = {Journal of the Atmospheric Sciences},
	author = {Davies-Jones, Robert},
	extradate = {2024-03-25},
	date = {1984-10-15},
    year = {1984},
	note = {Publisher: American Meteorological Society
Section: Journal of the Atmospheric Sciences},

}

@article{johnson_evaluation_2014,
	title = {Evaluation of Sounding-Derived Thermodynamic and Wind-Related Parameters Associated with Large Hail Events},
	volume = {9},
	rights = {https://creativecommons.org/licenses/by/4.0},
	issn = {1559-5404},
	extra = {https://ejssm.com/ojs/index.php/site/article/view/57},
	doi = {10.55599/ejssm.v9i5.57},
	pages = {1--42},
	number = {5},
	journal = {E-Journal of Severe Storms Meteorology},
	shortjournal = {{EJSSM}},
	author = {Johnson, Aaron W. and Sugden, Kelly E.},
	extradate = {2024-03-25},
	date = {2014-12-09},
    year = {2014},
	langid = {english},

}

@article {coffer_ealry_evening_transition_2015,
      author = "Brice E. Coffer and Matthew D. Parker",
      title = "Impacts of Increasing Low-Level Shear on Supercells during the Early Evening Transition",
      journal = "Monthly Weather Review",
      year = "2015",
      publisher = "American Meteorological Society",
      address = "Boston MA, USA",
      volume = "143",
      number = "5",
      doi = "10.1175/MWR-D-14-00328.1",
      pages="1945--1969",
      extra = "https://journals.ametsoc.org/view/journals/mwre/143/5/mwr-d-14-00328.1.xml"
}

@article { brown_EarlyEveningTransition_2021,
      author = "Matthew C. Brown and Christopher J. Nowotarski and Andrew R. Dean and Bryan T. Smith and Richard L. Thompson and John M. Peters",
      title = "The Early Evening Transition in Southeastern {U.S.} Tornado Environments",
      journal = "Weather and Forecasting",
      year = "2021",
      publisher = "American Meteorological Society",
      address = "Boston MA, USA",
      volume = "36",
      number = "4",
      doi = "10.1175/WAF-D-20-0191.1",
      pages="1431--1452",
      extra = "https://journals.ametsoc.org/view/journals/wefo/36/4/WAF-D-20-0191.1.xml"
}

@article {gropp_NocturnalTransition_2018,
      author = "Matthew E. Gropp and Casey E. Davenport",
      title = "The Impact of the Nocturnal Transition on the Lifetime and Evolution of Supercell Thunderstorms in the {Great Plains}",
      journal = "Weather and Forecasting",
      year = "2018",
      publisher = "American Meteorological Society",
      address = "Boston MA, USA",
      volume = "33",
      number = "4",
      doi = "10.1175/WAF-D-17-0150.1",
      pages="1045--1061",
      extra = "https://journals.ametsoc.org/view/journals/wefo/33/4/waf-d-17-0150_1.xml"
}

@article{dennis_impact_2017,
	title = {The Impact of Vertical Wind Shear on Hail Growth in Simulated Supercells},
	volume = {74},
	issn = {0022-4928, 1520-0469},
	extra = {https://journals.ametsoc.org/view/journals/atsc/74/3/jas-d-16-0066.1.xml},
	doi = {10.1175/JAS-D-16-0066.1},

	pages = {641--663},
	number = {3},
	journal = {Journal of the Atmospheric Sciences},
	author = {Dennis, Eli J. and Kumjian, Matthew R.},
	extradate = {2024-09-05},
	date = {2017-03-01},
    year = {2017},
	note = {Publisher: American Meteorological Society
Section: Journal of the Atmospheric Sciences},

}

@article{lin_influences_2022,
	title = {Influences of {CAPE} on Hail Production in Simulated Supercell Storms},
	volume = {79},
	issn = {0022-4928, 1520-0469},
	extra = {https://journals.ametsoc.org/view/journals/atsc/79/1/JAS-D-21-0054.1.xml},
	doi = {10.1175/JAS-D-21-0054.1},
	pages = {179--204},
	number = {1},
	journal = {Journal of the Atmospheric Sciences},
	author = {Lin, Yuzhu and Kumjian, Matthew R.},
	extradate = {2024-09-07},
	date = {2022-01-07},
    year = {2022},
	note = {Publisher: American Meteorological Society
Section: Journal of the Atmospheric Sciences},

}

@article{thompson_effective_2007,
	title = {Effective Storm-Relative Helicity and Bulk Shear in Supercell Thunderstorm Environments},
	volume = {22},
	issn = {1520-0434, 0882-8156},
	extra = {https://journals.ametsoc.org/view/journals/wefo/22/1/waf969_1.xml},
	doi = {10.1175/WAF969.1},
	pages = {102--115},
	number = {1},
	journal = {Weather and Forecasting},
	author = {Thompson, Richard L. and Mead, Corey M. and Edwards, Roger},
	extradate = {2024-10-11},
	date = {2007-02-01},
    year = {2007},
	note = {Publisher: American Meteorological Society
Section: Weather and Forecasting},

}

@article{bunkers_documentation_2007,
	title = {Documentation of a Rare Tornadic Left-Moving Supercell},
	volume = {2},
	rights = {Copyright (c) 2007},
	issn = {1559-5404},
	extra = {https://ejssm.com/ojs/index.php/site/article/view/7},
	doi = {10.55599/ejssm.v2i2.7},
	pages = {1--22},
	number = {2},
	journal = {E-Journal of Severe Storms Meteorology},
	author = {Bunkers, Matthew J. and Stoppkotte, John W.},
	extradate = {2024-10-30},
	date = {2007-01-31},
    year = {2007},
	langid = {english},
	note = {Number: 2},

}

@article{monteverdi_first_2001,
	title = {First {WSR}-88{D} Documentation of an Anticyclonic Supercell with Anticyclonic Tornadoes: The {Sunnyvale–Los Altos, California}, Tornadoes of 4 {May} 1998},
	volume = {129},
	issn = {1520-0493, 0027-0644},
	extra = {https://journals.ametsoc.org/view/journals/mwre/129/11/1520-0493_2001_129_2805_fwdoaa_2.0.co_2.xml},
	doi = {10.1175/1520-0493(2001)129<2805:FWDOAA>2.0.CO;2},
	shorttitle = {First {WSR}-88D Documentation of an Anticyclonic Supercell with Anticyclonic Tornadoes},

	pages = {2805--2814},
	number = {11},
	journal = {Monthly Weather Review},
	author = {Monteverdi, John P. and Blier, Warren and Stumpf, Greg and Pi, Wilfred and Anderson, Karl},
	extradate = {2024-10-30},
	date = {2001-11-01},
    year = {2001},
	note = {Publisher: American Meteorological Society
Section: Monthly Weather Review},

}

@article{brown_iowa_1980,
	title = {The {Iowa} Cyclonic-Anticyclonic Tornado Pair and Its Parent Thunderstorm},
	volume = {108},
	issn = {1520-0493, 0027-0644},
	extra = {https://journals.ametsoc.org/view/journals/mwre/108/10/1520-0493_1980_108_1626_ticatp_2_0_co_2.xml},
	doi = {10.1175/1520-0493(1980)108<1626:TICATP>2.0.CO;2},
	pages = {1626--1646},
	number = {10},
	journal = {Monthly Weather Review},
	author = {Brown, John M. and Knupp, Kevin R.},
	extradate = {2024-10-30},
	date = {1980-10-01},
    year = {1980},
	note = {Publisher: American Meteorological Society
Section: Monthly Weather Review},

}

@article{bunkers_predicting_2000,
	title = {Predicting Supercell Motion Using a New Hodograph Technique},
	volume = {15},
	issn = {1520-0434, 0882-8156},
	extra = {https://journals.ametsoc.org/view/journals/wefo/15/1/1520-0434_2000_015_0061_psmuan_2_0_co_2.xml},
	doi = {10.1175/1520-0434(2000)015<0061:PSMUAN>2.0.CO;2},
	pages = {61--79},
	number = {1},
	journal = {Weather and Forecasting},
	author = {Bunkers, Matthew J. and Klimowski, Brian A. and Zeitler, Jon W. and Thompson, Richard L. and Weisman, Morris L.},
	extradate = {2024-10-30},
	date = {2000-02-01},
    year = {2000},
	note = {Publisher: American Meteorological Society
Section: Weather and Forecasting},

}

@article{edwards_photographic_2006,
	title = {Photographic Documentation and Environmental Analysis of an Intense, Anticyclonic Supercell on the {Colorado} Plains},
	volume = {134},
	issn = {1520-0493, 0027-0644},
	extra = {https://journals.ametsoc.org/view/journals/mwre/134/12/mwr3296.1.xml},
	doi = {10.1175/MWR3296.1},
	pages = {3753--3763},
	number = {12},
	journal = {Monthly Weather Review},
	author = {Edwards, Roger and Hodanish, Stephen J.},
	extradate = {2024-10-30},
	date = {2006-12-01},
    year = {2006},
	note = {Publisher: American Meteorological Society
Section: Monthly Weather Review},

}

@article{lindsey_observations_2005,
	title = {Observations of a Severe, Left-Moving Supercell on 4 {May} 2003},
	volume = {20},
	issn = {1520-0434, 0882-8156},
	extra = {https://journals.ametsoc.org/view/journals/wefo/20/1/waf-830_1.xml},
	doi = {10.1175/WAF-830.1},
	pages = {15--22},
	number = {1},
	journal = {Weather and Forecasting},
	author = {Lindsey, Daniel T. and Bunkers, Matthew J.},
	extradate = {2024-10-30},
	date = {2005-02-01},
    year = {2005},
	note = {Publisher: American Meteorological Society
Section: Weather and Forecasting},

}

@article{brown_evolution_1994,
	title = {Evolution and Morphology of Two Splitting Thunderstorms with Dominant Left-Moving Members},
	volume = {122},
	issn = {1520-0493, 0027-0644},
	extra = {https://journals.ametsoc.org/view/journals/mwre/122/9/1520-0493_1994_122_2052_eamots_2_0_co_2.xml},
	doi = {10.1175/1520-0493(1994)122<2052:EAMOTS>2.0.CO;2},
	pages = {2052--2067},
	number = {9},
	journal = {Monthly Weather Review},
	author = {Brown, Rodger A. and Meitín, Rebecca J.},
	extradate = {2024-10-30},
	date = {1994-09-01},
    year = {1994},
	note = {Publisher: American Meteorological Society
Section: Monthly Weather Review},

}

@article{nielsen-gammon_detection_1995,
	title = {Detection and Interpretation of Left-Moving Severe Thunderstorms Using the {WSR}-88{D}: A Case Study},
	volume = {10},
	issn = {1520-0434, 0882-8156},
	extra = {https://journals.ametsoc.org/view/journals/wefo/10/1/1520-0434_1995_010_0127_daiolm_2_0_co_2.xml},
	doi = {10.1175/1520-0434(1995)010<0127:DAIOLM>2.0.CO;2},
	shorttitle = {Detection and Interpretation of Left-Moving Severe Thunderstorms Using the {WSR}-88D},
	pages = {127--140},
	number = {1},
	journal = {Weather and Forecasting},
	author = {Nielsen-Gammon, John W. and Read, William L.},
	extradate = {2024-10-30},
	date = {1995-03-01},
    year = {1995},
	note = {Publisher: American Meteorological Society
Section: Weather and Forecasting},

}

@article{edwards_assessment_2004,
	title = {Assessment of anticyclonic supercell environments using close proximity soundings from the {RUC} model.},
	author = {Edwards, Roger and Thompson, Richard L and Mead, Corey M},
	year ={2004},
	langid = {english},
	journal = {22d Conf. on Severe Local Storms, Hyannis, MA, Amer. Meteor. Soc.},

}

@article{bunkers_vertical_2002,
	title = {Vertical Wind Shear Associated with Left-Moving Supercells},
	volume = {17},
	issn = {1520-0434, 0882-8156},
	extra = {https://journals.ametsoc.org/view/journals/wefo/17/4/1520-0434_2002_017_0845_vwsawl_2_0_co_2.xml},
	doi = {10.1175/1520-0434(2002)017<0845:VWSAWL>2.0.CO;2},
	pages = {845--855},
	number = {4},
	journal = {Weather and Forecasting},
	author = {Bunkers, Matthew J.},
	extradate = {2024-10-30},
	date = {2002-08-01},
	year ={2002},
	note = {Publisher: American Meteorological Society
Section: Weather and Forecasting},

}

@article{fujita_split_1968,
	title = {Split of a Thunderstorm into Anticyclonic and Cyclonic Storms and Their Motion as Determined from Numerical Model Experiments},
	volume = {25},
	issn = {0022-4928, 1520-0469},
	extra = {https://journals.ametsoc.org/view/journals/atsc/25/3/1520-0469_1968_025_0416_soatia_2_0_co_2.xml},
	doi = {10.1175/1520-0469(1968)025<0416:SOATIA>2.0.CO;2},
	pages = {416--439},
	number = {3},
	journal = {Journal of the Atmospheric Sciences},
	author = {Fujita, Tetsuya and Grandoso, Hector},
	extradate = {2024-10-30},
	date = {1968-05-01},
	year ={1968},
	note = {Publisher: American Meteorological Society
Section: Journal of the Atmospheric Sciences},

}

@article{dostalek_aspects_2004,
	title = {Aspects of a Tornadic Left-Moving Thunderstorm of 25 {May} 1999},
	volume = {19},
	issn = {1520-0434, 0882-8156},
	extra = {https://journals.ametsoc.org/view/journals/wefo/19/3/1520-0434_2004_019_0614_aoatlt_2_0_co_2.xml},
	doi = {10.1175/1520-0434(2004)019<0614:AOATLT>2.0.CO;2},
	pages = {614--626},
	number = {3},
	journal = {Weather and Forecasting},
	author = {Dostalek, John F. and Weaver, John F. and Phillips, G. Loren},
	extradate = {2024-10-30},
	date = {2004-06-01},
	year = {2004},
	note = {Publisher: American Meteorological Society
Section: Weather and Forecasting},

}

@article{goldacker_assessing_2023,
	title = {Assessing the Comparative Effects of Storm-Relative Helicity Components within Right-Moving Supercell Environments},
	volume = {80},
	issn = {0022-4928, 1520-0469},
	extra = {https://journals.ametsoc.org/view/journals/atsc/80/12/JAS-D-22-0253.1.xml},
	doi = {10.1175/JAS-D-22-0253.1},

	pages = {2805--2822},
	number = {12},
	journal = {Journal of the Atmospheric Sciences},
	author = {Goldacker, Nicholas A. and Parker, Matthew D.},
	extradate = {2024-10-30},
	date = {2023-12-05},
	year = {2023},
	note = {Publisher: American Meteorological Society
Section: Journal of the Atmospheric Sciences},
}

@article{bocheva_severe_2018,
	title = {Severe convective supercell outbreak over western {Bulgaria} on {July} 8, 2014},
	volume = {122},
	issn = {03246329},
	extra = {http://met.hu/ismeret-tar/kiadvanyok/idojaras/index.php?no=2018.2.5},
	doi = {10.28974/idojaras.2018.2.5},
	pages = {177--202},
	number = {2},
	journal = {Időjárás},
	shortjournal = {Időjárás},
	author = {Bocheva, Lilia and Dimitrova, Tsvetelina and Penchev, Rosen and Gospodinov, Ilian and Simeonov, Petio},
	extradate = {2024-11-01},
	year ={2018},
	langid = {english},

}

@article{houze_hailstorms_1993,
	title = {Hailstorms in {Switzerland}: Left Movers, Right Movers, and False Hooks},
	volume = {121},
	issn = {1520-0493, 0027-0644},
	extra = {https://journals.ametsoc.org/view/journals/mwre/121/12/1520-0493_1993_121_3345_hislmr_2_0_co_2.xml},
	doi = {10.1175/1520-0493(1993)121<3345:HISLMR>2.0.CO;2},
	shorttitle = {Hailstorms in Switzerland},
	pages = {3345--3370},
	number = {12},
	journal = {Monthly Weather Review},
	author = {Houze, R. A. and Schmid, W. and Fovell, R. G. and Schiesser, H.-H.},
	extradate = {2024-11-01},
	date = {1993-12-01},
	year = {1993},
	note = {Publisher: American Meteorological Society
Section: Monthly Weather Review},

}

@article{groenemeijer_sounding-derived_2007,
	title = {Sounding-derived parameters associated with large hail and tornadoes in the {Netherlands}},
	volume = {83},
	rights = {https://www.elsevier.com/tdm/userlicense/1.0/},
	issn = {01698095},
	extra = {https://linkinghub.elsevier.com/retrieve/pii/S0169809506001165},
	doi = {10.1016/j.atmosres.2005.08.006},
	pages = {473--487},
	number = {2},
	journal = {Atmospheric Research},
	shortjournal = {Atmospheric Research},
	author = {Groenemeijer, P.H. and Van Delden, A.},
	extradate = {2024-11-02},
	date = {2007-02},
	year = {2007},
	langid = {english},

}

@article{davies-jones_review_2015,
	title = {A review of supercell and tornado dynamics},
	volume = {158-159},
	issn = {01698095},
	extra = {https://linkinghub.elsevier.com/retrieve/pii/S0169809514001756},
	doi = {10.1016/j.atmosres.2014.04.007},
	pages = {274--291},
	journal = {Atmospheric Research},
	shortjournal = {Atmospheric Research},
	author = {Davies-Jones, Robert},
	extradate = {2024-11-02},
	date = {2015-05},
	year = {2015},
	langid = {english},

}

@article{grasso_dissipation_2000,
	title = {The Dissipation of a Left-Moving Cell in a Severe Storm Environment},
	volume = {128},
	issn = {1520-0493, 0027-0644},
	extra = {https://journals.ametsoc.org/view/journals/mwre/128/8/1520-0493_2000_128_2797_tdoalm_2.0.co_2.xml},
	doi = {10.1175/1520-0493(2000)128<2797:TDOALM>2.0.CO;2},
	pages = {2797--2815},
	number = {8},
	journal = {Monthly Weather Review},
	author = {Grasso, Lewis D.},
	extradate = {2024-11-02},
	date = {2000-08-01},
	year = {2000},
	note = {Publisher: American Meteorological Society
Section: Monthly Weather Review},
}

@article{grasso_observations_2001,
	title = {Observations of a Severe Left Moving Thunderstorm},
	volume = {16},
	issn = {1520-0434, 0882-8156},
	extra = {https://journals.ametsoc.org/view/journals/wefo/16/4/1520-0434_2001_016_0500_ooaslm_2_0_co_2.xml},
	doi = {10.1175/1520-0434(2001)016<0500:OOASLM>2.0.CO;2},
	pages = {500--511},
	number = {4},
	journal = {Weather and Forecasting},
	author = {Grasso, Lewis D. and Hilgendorf, Eric R.},
	extradate = {2024-11-02},
	date = {2001-08-01},
	year = {2001},
	note = {Publisher: American Meteorological Society
Section: Weather and Forecasting},
}

@article{smith_convective_2012,
	title = {Convective Modes for Significant Severe Thunderstorms in the Contiguous {United States}. {Part} {I}: Storm Classification and Climatology},
	volume = {27},
	issn = {1520-0434, 0882-8156},
	extra = {https://journals.ametsoc.org/view/journals/wefo/27/5/waf-d-11-00115_1.xml},
	doi = {10.1175/WAF-D-11-00115.1},
	shorttitle = {Convective Modes for Significant Severe Thunderstorms in the Contiguous {United States}. {Part} I},
	pages = {1114--1135},
	number = {5},
	journal = {Weather and Forecasting},
	author = {Smith, Bryan T. and Thompson, Richard L. and Grams, Jeremy S. and Broyles, Chris and Brooks, Harold E.},
	extradate = {2024-11-02},
	date = {2012-10-01},
	year = {2012},
	note = {Publisher: American Meteorological Society
Section: Weather and Forecasting},
}

@article{tonn_evaluating_2023,
	title = {Evaluating {Bunkers}’ storm motion of hail-producing supercells and their storm-relative helicity in {Germany}},
	volume = {32},
	pages = {229--243},
	author = {Tonn, Mathis and Wilhelm, Jannik and Kunz, Michael},
	date = {2023-08-03},
	year ={2023},
    language = {en},
    journal = {Meteorologische Zeitschrift},
    DOI = {10.1127/metz/2023/1165},
}

@article{charba_structure_1971,
	title = {Structure and Movement of the Severe Thunderstorms of 3 {April} 1964 as Revealed from Radar and Surface Mesonetwork Data Analysis},
	volume = {49},
	doi = {10.2151/jmsj1965.49.3_191},
	pages = {191--214},
	number = {3},
	journal = {Journal of the Meteorological Society of Japan. Ser. {II}},
	author = {Charba, Jess and Sasaki, Yoshikazu},
	year ={1971},
}

@article{bunkers_observational_2006-1,
	title = {An Observational Examination of Long-Lived Supercells. {Part} {II}: Environmental Conditions and Forecasting},
	volume = {21},
	issn = {1520-0434, 0882-8156},
	extra = {https://journals.ametsoc.org/view/journals/wefo/21/5/waf952_1.xml},
	doi = {10.1175/WAF952.1},
	shorttitle = {An Observational Examination of Long-Lived Supercells. {Part} {II}},
	pages = {689--714},
	number = {5},
	journal = {Weather and Forecasting},
	author = {Bunkers, Matthew J. and Johnson, Jeffrey S. and Czepyha, Lee J. and Grzywacz, Jason M. and Klimowski, Brian A. and Hjelmfelt, Mark R.},
	extradate = {2024-11-03},
	date = {2006-10-01},
    year = {2006},
	note = {Publisher: American Meteorological Society
Section: Weather and Forecasting},
}

@book{hammond_study_1967,
	location = {Norman, Oklahoma},
	title = {Study of a left moving thunderstorm of 23 {April} 1964},
	series = {{ESSA} technical memorandum. {IERTM}-{NSSL}},
	number = {31},
	publisher = {U.S. Department of Commerce, Environmental Science Services Administration, Institutes for Environmental Research, National Severe Storms Laboratory},
	author = {Hammond, George R.},
	editora = {Harrold, Terence W. and Burnham, Jack and Spavins, Clifford S. and {National Severe Storms Laboratory}},
	editoratype = {collaborator},
	year ={1967},
    pages = {1--75},
}

@article{brimelow_modeling_2002,
	title = {Modeling Maximum Hail Size in {Alberta} Thunderstorms},
	volume = {17},
	issn = {1520-0434, 0882-8156},
	extra = {https://journals.ametsoc.org/view/journals/wefo/17/5/1520-0434_2002_017_1048_mmhsia_2_0_co_2.xml},
	doi = {10.1175/1520-0434(2002)017<1048:MMHSIA>2.0.CO;2},
	pages = {1048--1062},
	number = {5},
	journal = {Weather and Forecasting},
	author = {Brimelow, Julian C. and Reuter, Gerhard W. and Poolman, Eugene R.},
	extradate = {2024-11-15},
	date = {2002-10-01},
	year = {2002},
	note = {Publisher: American Meteorological Society
Section: Weather and Forecasting},
}

@article{edwards_nationwide_1998,
	title = {Nationwide Comparisons of Hail Size with {WSR}-88{D} Vertically Integrated Liquid Water and Derived Thermodynamic Sounding Data},
	volume = {13},
	issn = {1520-0434, 0882-8156},
	extra = {https://journals.ametsoc.org/view/journals/wefo/13/2/1520-0434_1998_013_0277_ncohsw_2_0_co_2.xml},
	doi = {10.1175/1520-0434(1998)013<0277:NCOHSW>2.0.CO;2},
	pages = {277--285},
	number = {2},
	journal = {Weather and Forecasting},
	author = {Edwards, Roger and Thompson, Richard L.},
	extradate = {2024-11-15},
	date = {1998-06-01},
	year = {1998},
	note = {Publisher: American Meteorological Society
Section: Weather and Forecasting},
}

@article{pucik_proximity_2015,
	title = {Proximity Soundings of Severe and Nonsevere Thunderstorms in Central {Europe}},
	volume = {143},
	issn = {1520-0493, 0027-0644},
	extra = {https://journals.ametsoc.org/view/journals/mwre/143/12/mwr-d-15-0104.1.xml},
	doi = {10.1175/MWR-D-15-0104.1},

	pages = {4805--4821},
	number = {12},
	journal = {Monthly Weather Review},
	author = {Púčik, Tomáš and Groenemeijer, Pieter and Rýva, David and Kolář, Miroslav},
	extradate = {2024-11-15},
	date = {2015-12-01},
	year = {2015},
	note = {Publisher: American Meteorological Society
Section: Monthly Weather Review},
}

@article{giordani_characterizing_2024,
	title = {Characterizing hail-prone environments using convection-permitting reanalysis and overshooting top detections over south-central {Europe}},
	volume = {24},
	issn = {1561-8633},
	extra = {https://nhess.copernicus.org/articles/24/2331/2024/},
	doi = {10.5194/nhess-24-2331-2024},
	pages = {2331--2357},
	number = {7},
	journal = {Natural Hazards and Earth System Sciences},
	author = {Giordani, Antonio and Kunz, Michael and Bedka, Kristopher M. and Punge, Heinz Jürgen and Paccagnella, Tiziana and Pavan, Valentina and Cerenzia, Ines M. L. and Di Sabatino, Silvana},
	extradate = {2024-11-18},
	date = {2024-07-12},
	year = {2024},
	note = {Publisher: Copernicus {GmbH}},
}

@article{pounds_analysis_2024,
	title = {Analysis of Hail Production via Simulated Hailstone Trajectories in the 29 {May} 2012 {Kingfisher}, {Oklahoma}, Supercell},
	volume = {152},
	issn = {1520-0493, 0027-0644},
	extra = {https://journals.ametsoc.org/view/journals/mwre/152/1/MWR-D-23-0073.1.xml},
	doi = {10.1175/MWR-D-23-0073.1},
	pages = {245--276},
	number = {1},
	journal = {Monthly Weather Review},
	author = {Pounds, Lauren E. and Ziegler, Conrad L. and Adams-Selin, Rebecca D. and Biggerstaff, Michael I.},
	extradate = {2024-11-27},
	date = {2024-01-02},
	year = {2024},
	note = {Publisher: American Meteorological Society
Section: Monthly Weather Review},
}

@article{zeitler_operational_2005,
	title = {Operational Forecasting of Supercell Motion: Review and Case Studies Using Multiple Datasets},
	volume = {29},
	shorttitle = {Operational Forecasting of Supercell Motion},
	pages = {81--97},
	journal = {National Weather Digest},
	shortjournal = {National Weather Digest},
	author = {Zeitler, Jon and Bunkers, Matthew},
	date = {2005-01-01},
	year = {2005},
}

@incollection{zipser_views_2003,
	location = {Boston, {MA}},
	title = {Some Views On “Hot Towers” after 50 Years of Tropical Field Programs and Two Years of {TRMM} Data},
	isbn = {978-1-878220-63-9},
	extra = {https://doi.org/10.1007/978-1-878220-63-9_5},
	pages = {49--58},
	booktitle = {Cloud Systems, Hurricanes, and the Tropical Rainfall Measuring Mission ({TRMM}): A Tribute to Dr. Joanne Simpson},
	publisher = {American Meteorological Society},
	author = {Zipser, Edward J.},
	editor = {Tao, Wei-Kuo and Adler, Robert},
	extradate = {2024-12-16},
	year ={2003},
	langid = {english},
	doi = {10.1007/978-1-878220-63-9_5},
}

@article{romps_undiluted_2010,
	title = {Do Undiluted Convective Plumes Exist in the Upper Tropical Troposphere?},
	volume = {67},
	issn = {0022-4928, 1520-0469},
	extra = {https://journals.ametsoc.org/view/journals/atsc/67/2/2009jas3184.1.xml},
	doi = {10.1175/2009JAS3184.1},
	pages = {468--484},
	number = {2},
	journal = {Journal of the Atmospheric Sciences},
	author = {Romps, David M. and Kuang, Zhiming},
	extradate = {2024-12-16},
	date = {2010-02-01},
	year = {2010},
	note = {Publisher: American Meteorological Society
Section: Journal of the Atmospheric Sciences},
}

@article{romps_nature_2010,
	title = {Nature versus Nurture in Shallow Convection},
	volume = {67},
	issn = {0022-4928, 1520-0469},
	extra = {https://journals.ametsoc.org/view/journals/atsc/67/5/2009jas3307.1.xml},
	doi = {10.1175/2009JAS3307.1},
	pages = {1655--1666},
	number = {5},
	journal = {Journal of the Atmospheric Sciences},
	author = {Romps, David M. and Kuang, Zhiming},
	extradate = {2024-12-16},
	date = {2010-05-01},
	year = {2010},
	note = {Publisher: American Meteorological Society
Section: Journal of the Atmospheric Sciences},
}

@article{zeeb_supercell_2024,
	title = {Supercell precipitation contribution to the {United States} hydroclimate},
	volume = {44},
	rights = {© 2024 Royal Meteorological Society},
	issn = {1097-0088},
	extra = {https://onlinelibrary.wiley.com/doi/abs/10.1002/joc.8395},
	doi = {10.1002/joc.8395},
	pages = {1489--1512},
	number = {5},
	journal = {International Journal of Climatology},
	author = {Zeeb, Aaron W. and Ashley, Walker S. and Haberlie, Alex M. and Gensini, Vittorio A. and Michaelis, Allison C.},
	extradate = {2024-12-17},
	year ={2024},
	langid = {english},
	note = {\_eprint: https://onlinelibrary.wiley.com/doi/pdf/10.1002/joc.8395},
}

@article{ashley_future_2023,
	title = {The Future of Supercells in the {United States}},
	volume = {104},
	issn = {0003-0007, 1520-0477},
	extra = {https://journals.ametsoc.org/view/journals/bams/104/1/BAMS-D-22-0027.1.xml},
	doi = {10.1175/BAMS-D-22-0027.1},
	pages = {E1--E21},
	number = {1},
	journal = {Bulletin of the American Meteorological Society},
	author = {Ashley, Walker S. and Haberlie, Alex M. and Gensini, Vittorio A.},
	extradate = {2024-12-17},
	date = {2023-01-04},
	year = {2023},
	note = {Publisher: American Meteorological Society
Section: Bulletin of the American Meteorological Society},
}

@misc{gunturi_impact_2017,
	title = {Impact of {ENSO} on {U.S.} Tornado and Hail Frequencies},
    howpublished = {\url{https://www.columbia.edu/~mkt14/files/WillisRe_Impact_of_ENSO_on_US_Tornado_and_Hail_frequencies_Final.pdf}},
	author = {Gunturi, P. and Tippett, M.},
	extradate = {2024-12-18},
	date = {2017-03},
    year = {2017},
    journal = {Managing Severe Thunderstorm Risk Tech. Rep.},
	note = {page 5.},
    note = {Accessed 30-04-2025},
}

@article{wade_dynamics_2021,
	title = {Dynamics of Simulated High-Shear, Low-{CAPE} Supercells},
	volume = {78},
	issn = {0022-4928, 1520-0469},
	extra = {https://journals.ametsoc.org/view/journals/atsc/78/5/JAS-D-20-0117.1.xml},
	doi = {10.1175/JAS-D-20-0117.1},
	pages = {1389--1410},
	number = {5},
	journal = {Journal of the Atmospheric Sciences},
	author = {Wade, Andrew R. and Parker, Matthew D.},
	extradate = {2025-01-08},
	date = {2021-05-01},
    year = {2021},
	note = {Publisher: American Meteorological Society
Section: Journal of the Atmospheric Sciences},
}

@article{klemp_dynamics_1987,
	title = {Dynamics of Tornadic Thunderstorms},
	volume = {19},
	issn = {0066-4189, 1545-4479},
	extra = {https://www.annualreviews.org/content/journals/10.1146/annurev.fl.19.010187.002101},
	doi = {10.1146/annurev.fl.19.010187.002101},
	pages = {369--402},
	issue = {Volume 19, 1987},
	journal = {Annual Review of Fluid Mechanics},
	author = {Klemp, J. B.},
	extradate = {2025-01-17},
	date = {1987-01-01},
    year = {1987},
	langid = {english},
	note = {Publisher: Annual Reviews},
}

@article{davenport_environmental_2021,
	title = {Environmental Evolution of Long-Lived Supercell Thunderstorms in the {Great Plains}},
	volume = {36},
	issn = {1520-0434, 0882-8156},
	extra = {https://journals.ametsoc.org/view/journals/wefo/36/6/WAF-D-21-0042.1.xml},
	doi = {10.1175/WAF-D-21-0042.1},
	pages = {2187--2209},
	number = {6},
	journal = {Weather and Forecasting},
	author = {Davenport, Casey E.},
	extradate = {2025-01-23},
	date = {2021-12-01},
    year = {2021},
	note = {Publisher: American Meteorological Society
Section: Weather and Forecasting},
}

@article{fischer_supercell_2024,
	title = {Supercell Tornadogenesis: Recent Progress in Our State of Understanding},
	volume = {105},
	issn = {0003-0007, 1520-0477},
	extra = {https://journals.ametsoc.org/view/journals/bams/105/7/BAMS-D-23-0031.1.xml},
	doi = {10.1175/BAMS-D-23-0031.1},
	shorttitle = {Supercell Tornadogenesis},
	pages = {E1084--E1097},
	number = {7},
	journal = {Bulletin of the American Meteorological Society},
	author = {Fischer, Jannick and Dahl, Johannes M. L. and Coffer, Brice E. and Houser, Jana Lesak and Markowski, Paul M. and Parker, Matthew D. and Weiss, Christopher C. and Schueth, Alex},
	extradate = {2025-02-05},
	date = {2024-07-09},
    year = {2024},
	note = {Publisher: American Meteorological Society
Section: Bulletin of the American Meteorological Society},
}

@article{banacos_use_2005,
	title = {The Use of Moisture Flux Convergence in Forecasting Convective Initiation: Historical and Operational Perspectives},
	volume = {20},
	issn = {1520-0434, 0882-8156},
	extra = {https://journals.ametsoc.org/view/journals/wefo/20/3/waf858_1.xml},
	doi = {10.1175/WAF858.1},
	shorttitle = {The Use of Moisture Flux Convergence in Forecasting Convective Initiation},
	pages = {351--366},
	number = {3},
	journal = {Weather and Forecasting},
	author = {Banacos, Peter C. and Schultz, David M.},
	extradate = {2025-02-19},
	date = {2005-06-01},
    year = {2005},
	note = {Publisher: American Meteorological Society
Section: Weather and Forecasting},
}

@article{grant_elevated_1995,
	title = {Elevated cold-sector severe thunderstorms: A preliminary study},
	volume = {19},
    pages = {1--7},
	number = {4},
	journal = {National Weather Digest},
	author = {Grant, Bradford N},
	year ={1995},
	langid = {english},
}

@article{reif_20-year_2017,
	title = {A 20-Year Climatology of Nocturnal Convection Initiation over the Central and Southern {Great Plains} during the Warm Season},
	volume = {145},
	issn = {1520-0493, 0027-0644},
	extra = {https://journals.ametsoc.org/view/journals/mwre/145/5/mwr-d-16-0340.1.xml},
	doi = {10.1175/MWR-D-16-0340.1},
	pages = {1615--1639},
	number = {5},
	journal = {Monthly Weather Review},
	author = {Reif, Dylan W. and Bluestein, Howard B.},
	extradate = {2025-02-19},
	date = {2017-05-01},
    year = {2017},
	note = {Publisher: American Meteorological Society
Section: Monthly Weather Review},
}

@article{horgan_five-year_2007,
	title = {A Five-Year Climatology of Elevated Severe Convective Storms in the {United States} East of the {Rocky Mountains}},
	volume = {22},
	issn = {1520-0434, 0882-8156},
	extra = {https://journals.ametsoc.org/view/journals/wefo/22/5/waf1032_1.xml},
	doi = {10.1175/WAF1032.1},
	pages = {1031--1044},
	number = {5},
	journal = {Weather and Forecasting},
	author = {Horgan, Katherine L. and Schultz, David M. and Hales, John E. and Corfidi, Stephen F. and Johns, Robert H.},
	extradate = {2025-02-19},
	date = {2007-10-01},
    year = {2007},
	note = {Publisher: American Meteorological Society
Section: Weather and Forecasting},
}

@misc{van_den_broeke_left-moving_2024,
	title = {Left-moving Supercells with Polarimetric Data Available, 2011--2022},
	extra = {https://zenodo.org/records/10655321},
	doi = {10.5281/zenodo.10655321},
	publisher = {Zenodo},
	author = {{Van Den Broeke}, Matthew and {Van Den Broeke}, Cynthia and Kirby, Simon and Schweigert, Benjamin and Nelson, Raychel},
	extradate = {2025-02-21},
	date = {2024-02-16},
    year = {2024},
}

@misc{lepore_xcape_2021,
	title = {xcape},
	extra = {https://zenodo.org/records/5270332},
	version = {v0.1.4},
	publisher = {Zenodo},
	author = {Lepore, Chiara and Allen, John and Abernathey, Ryan},
	extradate = {2025-02-21},
	date = {2021-07-26},
    year = {2021},
	doi = {10.5281/zenodo.5270332},
}

@article{macintosh_6_2017,
	title = {The 6 {May} 2010 Elevated Supercell during {VORTEX}2},
	volume = {145},
	issn = {1520-0493, 0027-0644},
	extra = {https://journals.ametsoc.org/view/journals/mwre/145/7/mwr-d-16-0329.1.xml},
	doi = {10.1175/MWR-D-16-0329.1},
	pages = {2635--2657},
	number = {7},
	journal = {Monthly Weather Review},
	author = {{MacIntosh}, Christopher W. and Parker, Matthew D.},
	extradate = {2025-02-24},
	date = {2017-07-01},
    year = {2017},
	note = {Publisher: American Meteorological Society
Section: Monthly Weather Review},
}

@article{nowotarski_characteristics_2011,
	title = {The Characteristics of Numerically Simulated Supercell Storms Situated over Statically Stable Boundary Layers},
	volume = {139},
	issn = {1520-0493, 0027-0644},
	extra = {https://journals.ametsoc.org/view/journals/mwre/139/10/mwr-d-10-05087.1.xml},
	doi = {10.1175/MWR-D-10-05087.1},
	pages = {3139--3162},
	number = {10},
	journal = {Monthly Weather Review},
	author = {Nowotarski, Christopher J. and Markowski, Paul M. and Richardson, Yvette P.},
	extradate = {2025-02-24},
	date = {2011-10-01},
    year = {2011},
	note = {Publisher: American Meteorological Society
Section: Monthly Weather Review},
}

@conference{andra_1997,
    author = {Andra, D. L., Jr.},
    booktitle = {28th Conf. on Radar Meteorology Amer. Meteor. Soc.},
    address = {Austin, TX},
    title = {The origin and evolution of the {WSR}-88{D} mesocyclone recognition nomogram},
    year = {1997},
    pages = {364--365}
}

@article{davies-jones_linear_nonlinear_propagation_2002,
    author = {Davies-Jones, Robert},
    date = {2002/11/01},
    language = {EN},
    extra	= {https://journals.ametsoc.org/view/journals/atsc/59/22/1520-0469_2003_059_3178_lanpos_2.0.co_2.xml},
 	publisher = {American Meteorological Society Section: Journal of the Atmospheric Sciences},
    volume = {59},
    pages = {3178--3205},
    journal = {Journal of the Atmospheric Sciences},
    doi = {10.1175/1520-0469(2003)059<3178:LANPOS>2.0.CO;2},
    issue =	{22},
    issn = {0022-4928, 1520-0469},
    title = {Linear and Nonlinear Propagation of Supercell Storms},
    year = {2002}
}

@article {Bunkers_2024_motion,
      author = "Matthew J. Bunkers and Matthew S. {Van Den Broeke} and John T. Allen",
      title = "An Update for Predicting Left-Moving Supercell Motion",
      journal = "Weather and Forecasting",
      year = "2024",
      publisher = "American Meteorological Society",
      address = "Boston MA, USA",
      volume = "39",
      number = "12",
      doi = "10.1175/WAF-D-24-0028.1",
      pages= "1777--1794",
      extra = "https://journals.ametsoc.org/view/journals/wefo/39/12/WAF-D-24-0028.1.xml"
}

@inproceedings{bothwell_2002,
    title={An integrated three-dimensional objective analysis scheme in use at the {Storm Prediction Center}},
    author={Bothwell, P. D. and J. A. Hart and R. L. Thompson},
    booktitle={21st Conf. on Severe Local Storms, San Antonio, TX},
    pages={JP3.1},
    year={2002},
    extra={https://ams.confex.com/ams/pdfpapers/47482.pdf}
}

@article {coniglio_2022,
      author = "Michael C. Coniglio and Ryan E. Jewell",
      title = "{SPC} Mesoscale Analysis Compared to Field-Project Soundings: Implications for Supercell Environment Studies",
      journal = "Monthly Weather Review",
      year = "2022",
      publisher = "American Meteorological Society",
      address = "Boston MA, USA",
      volume = "150",
      number = "3",
      doi = "10.1175/MWR-D-21-0222.1",
      pages= "567--588",
      extra = "https://journals.ametsoc.org/view/journals/mwre/150/3/MWR-D-21-0222.1.xml"
}

@article {coffer_2020,
      author = "Brice E. Coffer and Mateusz Taszarek and Matthew D. Parker",
      title = "Near-Ground Wind Profiles of Tornadic and Nontornadic Environments in the {United States} and {Europe} from {ERA5} Reanalyses",
      journal = "Weather and Forecasting",
      year = "2020",
      publisher = "American Meteorological Society",
      address = "Boston MA, USA",
      volume = "35",
      number = "6",
      doi = "10.1175/WAF-D-20-0153.1",
      pages= "2621--2638",
      extra = "https://journals.ametsoc.org/view/journals/wefo/35/6/waf-d-20-0153.1.xml"
}

@article { Benjamin_2004,
      author = "Stanley G. Benjamin and Dezsö Dévényi and Stephen S. Weygandt and Kevin J. Brundage and John M. Brown and Georg A. Grell and Dongsoo Kim and Barry E. Schwartz and Tatiana G. Smirnova and Tracy Lorraine Smith and Geoffrey S. Manikin",
      title = "An Hoextray Assimilation–Forecast Cycle: The {RUC}",
      journal = "Monthly Weather Review",
      year = "2004",
      publisher = "American Meteorological Society",
      address = "Boston MA, USA",
      volume = "132",
      number = "2",
      doi = "10.1175/1520-0493(2004)132%3C0495:AHACTR%3E2.0.CO;2",
      pages= "495--518",
      extra = "https://journals.ametsoc.org/view/journals/mwre/132/2/1520-0493_2004_132_0495_ahactr_2.0.co_2.xml"
}

@mastersthesis{brimelow_1999,
    author = {Brimelow, J. C.},
    title = {Numerical modelling of hailstone growth in {Alberta} storms},
    school = {University of Alberta,},
    department = {Earth and Atmospheric Sciences},
    year = {1999},
    page = {153},
    
}

@article {sherburn_2014,
      author = "Keith D.  Sherburn and Matthew D.  Parker",
      title = "Climatology and Ingredients of Significant Severe Convection in High-Shear, Low-{CAPE} Environments",
      journal = "Weather and Forecasting",
      year = "2014",
      publisher = "American Meteorological Society",
      address = "Boston MA, USA",
      volume = "29",
      number = "4",
      doi = "10.1175/WAF-D-13-00041.1",
      pages= "854--877",
      extra = "https://journals.ametsoc.org/view/journals/wefo/29/4/waf-d-13-00041_1.xml"
}

@article {sherburn_2016,
      author = "Keith D.  Sherburn and Matthew D.  Parker and Jessica R.  King and Gary M.  Lackmann",
      title = "Composite Environments of Severe and Nonsevere High-Shear, Low-{CAPE} Convective Events",
      journal = "Weather and Forecasting",
      year = "2016",
      publisher = "American Meteorological Society",
      address = "Boston MA, USA",
      volume = "31",
      number = "6",
      doi = "10.1175/WAF-D-16-0086.1",
      pages= "1899--1927",
      extra = "https://journals.ametsoc.org/view/journals/wefo/31/6/waf-d-16-0086_1.xml"
}

@conference{burgess1981evidence,
    title={Evidence for anticyclonic rotation in left-moving thunderstorms},
    author={Burgess, Donald W},
    booktitle={20th Conf. on Radar Meteorology, Boston, MA},
    pages={52--54},
    year={1981}
}

\end{document}